\documentclass[twocolumn,tighten,twocolappendix]{aastex63}
\pdfoutput=1 
\usepackage{amsmath,amstext}
\usepackage[T1]{fontenc}
\usepackage{ae,aecompl}
\usepackage[utf8]{inputenc}
\usepackage{apjfonts} 
\usepackage[figure,figure*]{hypcap}
\usepackage{enumitem}
\usepackage{bm}
\usepackage{pifont}
\input{hyperlink-year-only-natbib-patch}

\newcommand*{\https}[1]{\href{https://#1}{#1}}

\graphicspath{{./}{}}

\shorttitle{SIDM Signatures in MW--mass Halos}
\shortauthors{Nadler et al.}

\begin{document}

\title{Signatures of Velocity-Dependent Dark Matter Self-Interactions in Milky Way--mass Halos}

\correspondingauthor{Ethan~O.~Nadler}
\email{enadler@stanford.edu}
\author[0000-0002-1182-3825]{Ethan~O.~Nadler}
\affiliation{Kavli Institute for Particle Astrophysics and Cosmology and Department of Physics, Stanford University, Stanford, CA 94305, USA}
\author[0000-0002-5209-1173]{Arka~Banerjee}
\affiliation{Kavli Institute for Particle Astrophysics and Cosmology and Department of Physics, Stanford University, Stanford, CA 94305, USA}
\affiliation{SLAC National Accelerator Laboratory, Menlo Park, CA 94025, USA}
\author[0000-0002-0298-4432]{Susmita~Adhikari}
\affiliation{Kavli Institute for Particle Astrophysics and Cosmology and Department of Physics, Stanford University, Stanford, CA 94305, USA}
\author[0000-0002-1200-0820]{Yao-Yuan~Mao}
\altaffiliation{NHFP Einstein Fellow}
\affiliation{Department of Physics and Astronomy, Rutgers, The State University of New Jersey, Piscataway, NJ 08854, USA}
\author[0000-0003-2229-011X]{Risa~H.~Wechsler}
\affiliation{Kavli Institute for Particle Astrophysics and Cosmology and Department of Physics, Stanford University, Stanford, CA 94305, USA}
\affiliation{SLAC National Accelerator Laboratory, Menlo Park, CA 94025, USA}

\begin{abstract}
We explore the impact of elastic, anisotropic, velocity-dependent dark matter (DM) self-interactions on the host halo and subhalos of Milky Way (MW)--mass systems. We consider a generic self-interacting dark matter (SIDM) model parameterized by the masses of a light mediator and the DM particle. The ratio of these masses,~$w$, sets the velocity scale above which momentum transfer due to DM self-interactions becomes inefficient. We perform high-resolution zoom-in simulations of an MW-mass halo for values of $w$ that span scenarios in which self-interactions either between the host and its subhalos or only within subhalos efficiently transfer momentum, and we study the effects of self-interactions on the host halo and on the abundance, radial distribution, orbital dynamics, and density profiles of subhalos in each case. The abundance and properties of surviving subhalos are consistent with being determined primarily by subhalo--host halo interactions. In particular, subhalos on radial orbits in models with larger values of the cross section at the host halo velocity scale are more susceptible to tidal disruption owing to mass loss from ram pressure stripping caused by self-interactions with the host. This mechanism suppresses the abundance of surviving subhalos relative to collisionless DM simulations, with stronger suppression for larger values of $w$. Thus, probes of subhalo abundance around MW-mass hosts can be used to place upper limits on the self-interaction cross section at velocity scales of $\sim 200\ \rm{km\ s}^{-1}$, and combining these measurements with the orbital properties and internal dynamics of subhalos may break degeneracies among velocity-dependent SIDM models.
\end{abstract}

\keywords{\href{http://astrothesaurus.org/uat/353}{Dark matter (353)}; \href{http://astrothesaurus.org/uat/1049}{Milky Way dark matter halo (1049)}; \href{http://astrothesaurus.org/uat/574}{Galaxy abundances (574)}; \href{http://astrothesaurus.org/uat/1965}{Computational methods (1965)}}

\section{Introduction} \label{sec:intro}

Self-interacting dark matter (SIDM) has long been an attractive alternative to purely cold, collisionless dark matter (CDM) owing to several potential ``small-scale problems'' attributed to CDM (see \citealt{Bullock170704256} for a review of these problems, and see \citealt{Tulin170502358} for a review of their potential resolutions in SIDM). Historically, SIDM was motivated by the core--cusp problem, which concerns the apparent discrepancy between the steep, cuspy NFW profiles ubiquitous among DM halos in CDM simulations and the flatter, cored profiles inferred from the dynamics of various tracers in dwarf galaxies \citep{Firmani0002376,Spergel9909386,Dave0006218,Colin0205322}. The core--cusp problem is sensitive to the impact of baryonic feedback, and it remains an active area of study (e.g., \citealt{Elbert14121477}). In particular, it has recently been cast in terms of the diversity of inner DM density profiles for field \citep{Oman150401437,KamadaPrL,Kaplinghat19110900544,Ren180805695,Zavala190409998,SantosSantos191109116} or satellite \citep{Valli171103502,Kahlhoefer190410539} galaxies at fixed halo properties. For Milky Way (MW) satellite galaxies, self-interactions in the presence of the tidal field of the Galactic disk may accelerate gravothermal core collapse, and this process has been proposed as a unique signature of SIDM (\citealt{Kaplinghat190404939,Nishikawa190100499,Sameie190407872}).

Self-interactions can also affect the abundance of DM substructure in a statistical fashion. In particular, many authors have studied the subhalo populations of MW-mass hosts in the context of SIDM (e.g., \citealt{Dave0006218,Colin0205322,DOnghia0206125,Vogelsberger12015892,Rocha12083025,Zavala12116426,Dooley160308919,Vogelsberger180503203,Robles190301469}), and a few cosmological SIDM-plus-hydrodynamic simulations of isolated dwarfs \citep{Fry150100497,Harvey180810451,Fitts181111791} and MW-mass systems \citep{Robles170607514} have been performed.

The consensus of these studies has been that SIDM models that do not drastically reduce the number counts of surviving subhalos have little effect on subhalo abundances in MW-mass systems (see, e.g., the discussion in \citealt{Tulin170502358}). However, this is a model-dependent statement that warrants scrutiny in light of increasingly precise constraints on the abundance of subhalos in MW-mass systems enabled by recent measurements of the luminosity function of satellite galaxies around the MW (e.g., \citealt*{PaperI}) and around MW analogs (e.g., \citealt{Geha170506743,Kondapally180809020,Bennet190603230}). In addition, theoretical predictions for gaps and perturbations in nearby stellar streams (e.g., \citealt{Banik191102662,Bonaca181103631}), direct and indirect detection experiments (e.g., \citealt{Ibarra190800747,Ishiyama190703642}), and flux ratios in strongly lensed systems (e.g., \citealt{Gilman190806983,Hsueh190504182}) are all sensitive to the abundance and internal properties of DM substructure in distinct ways.

Many of the aforementioned SIDM studies only consider isotropic, velocity-independent self-interactions, at least on the velocity scales relevant for the MW host halo and its subhalos. Nonetheless, there are theoretical and observational arguments in favor of a velocity-\emph{dependent} SIDM cross section, which is a generic consequence of interactions in a nonminimal dark sector (see, e.g., \citealt{Feng09053039,Loeb10116374,Tulin13023898,Boddy14023629,Kaplinghat158003339,Tulin170502358}). While velocity-dependent SIDM models have been explored on galaxy cluster scales (e.g., \citealt{Robertson161203906,Banerjee190612026}), a thorough study of the corresponding predictions on Galactic scales is now crucial. This is particularly relevant because the increasingly well-measured abundance and internal properties of MW satellites can be affected by self-interactions at both the host halo velocity scale (which depend on satellites' typical velocities relative to the host) and subhalo velocity scales (which depend on the internal velocity dispersion of individual subhalos).

Herein we investigate the effects of an SIDM model with a generic, velocity-dependent self-interaction cross section on the DM host halo and subhalos of MW-mass systems. We focus on the physical mechanisms that shape the abundance and properties of surviving and disrupted subhalos, setting the stage for future analyses to constrain the SIDM cross section at the velocity scale of MW-mass systems. This paper is organized as follows: We describe our SIDM model in Section~\ref{sec:model} and our simulations in Section~\ref{sec:simulations}. We then present our results, focusing on the properties of our host halos in Section \ref{subsec:host} and their subhalo populations in Section \ref{subsec:sub}. We discuss challenges and caveats in Section \ref{sec:caveats}, we compare to previous studies in Section \ref{subsec:comparison}, we consider prospects for SIDM constraints in Section \ref{subsec:prospects}, and we conclude in Section \ref{sec:conclusion}.

\section{SIDM Model}
\label{sec:model}

We consider an SIDM model in which a DM particle $\chi$ interacts under the exchange of a light mediator, which can be either a scalar particle $\phi$ or a vector particle $\phi^\mu$. These scenarios are respectively described by the interaction Lagrangians \citep{Tulin170502358}
\begin{align}
\mathcal{L}_{\rm{int}} =
\left\{\begin{array}{ll}
    g_\chi \bar{\chi} \chi \phi,\ \rm{scalar\ mediator}\\
    g_\chi \bar{\chi} \gamma^\mu \chi \phi_\mu,\ \rm{vector\ mediator},\end{array}\right.\label{eq:lagrangian}
\end{align}
where $g_\chi$ is a coupling constant and $\gamma^\mu$ are Dirac matrices.

Self-interactions governed by Equation \ref{eq:lagrangian} can be described by a Yukawa potential
\begin{equation}
    V(r) = \pm \frac{\alpha_\chi}{r}e^{-m_{\phi}/r},
\end{equation}
where $r$ is the separation between DM particles, $\alpha_\chi\equiv g_\chi^2/4\pi$ is the analog of the fine-structure constant in the dark sector, and $m_{\phi}$ is the mediator mass.

We focus on $t$-channel scattering, which leads to an effective differential scattering cross section \citep{Ibe09125425,Kummer170604794}
\begin{equation}
    \frac{\text{d}\sigma}{\text{d}\Omega} = \frac{\sigma_0}{2\Big[1+\frac{v^2}{w^2}\sin^2\left(\frac{\theta}{2}\right)\Big]^2},\label{eq:xsec}
\end{equation}
where $v$ is the relative velocity between interacting DM particles with mass $m_\chi$, $w\equiv m_{\phi}/m_\chi$ is a characteristic velocity scale, $\sigma_0 \equiv 4\pi\alpha_\chi^2 m_{\chi}^2/m_{\phi}^4$ is the amplitude of the cross section, and $\theta$ is the scattering angle in the center-of-mass frame. For $w\ll v$ (in natural units), typically corresponding to an $\rm{MeV}$-scale mediator for a $\rm{GeV}$-scale DM particle mass, Equation \ref{eq:xsec} reduces to Rutherford-like scattering; for a heavy mediator, it reduces to velocity-independent isotropic scattering.

The corresponding momentum transfer cross section for identical particles is given by \citep{Kahlhoefer13083419}
\begin{equation}
    \sigma_T = \int \frac{\text{d}\sigma}{\text{d}\Omega} \left(1-\left |\cos\theta \right|\right)\rm{d}\Omega.\label{eq:sigma_T}
\end{equation}
Note that $\sigma_T$ is a function of $v$, $w$, and $\sigma_0$. Finally, the \emph{total} cross section
\begin{equation}
    \sigma = \int \frac{\text{d}\sigma}{\text{d}\Omega}\rm{d}\Omega\label{eq:sigma_tot}
\end{equation}
determines the probability of DM self-interactions in our numerical implementation (see Section \ref{sec:implementation}). For the case of isotropic, velocity-independent self-interactions, the total cross section is related to $\sigma_T$ and $\sigma_0$ via $\sigma = 2\sigma_T = 2\pi\sigma_0$.

\begin{figure}
\includegraphics[scale=0.4]{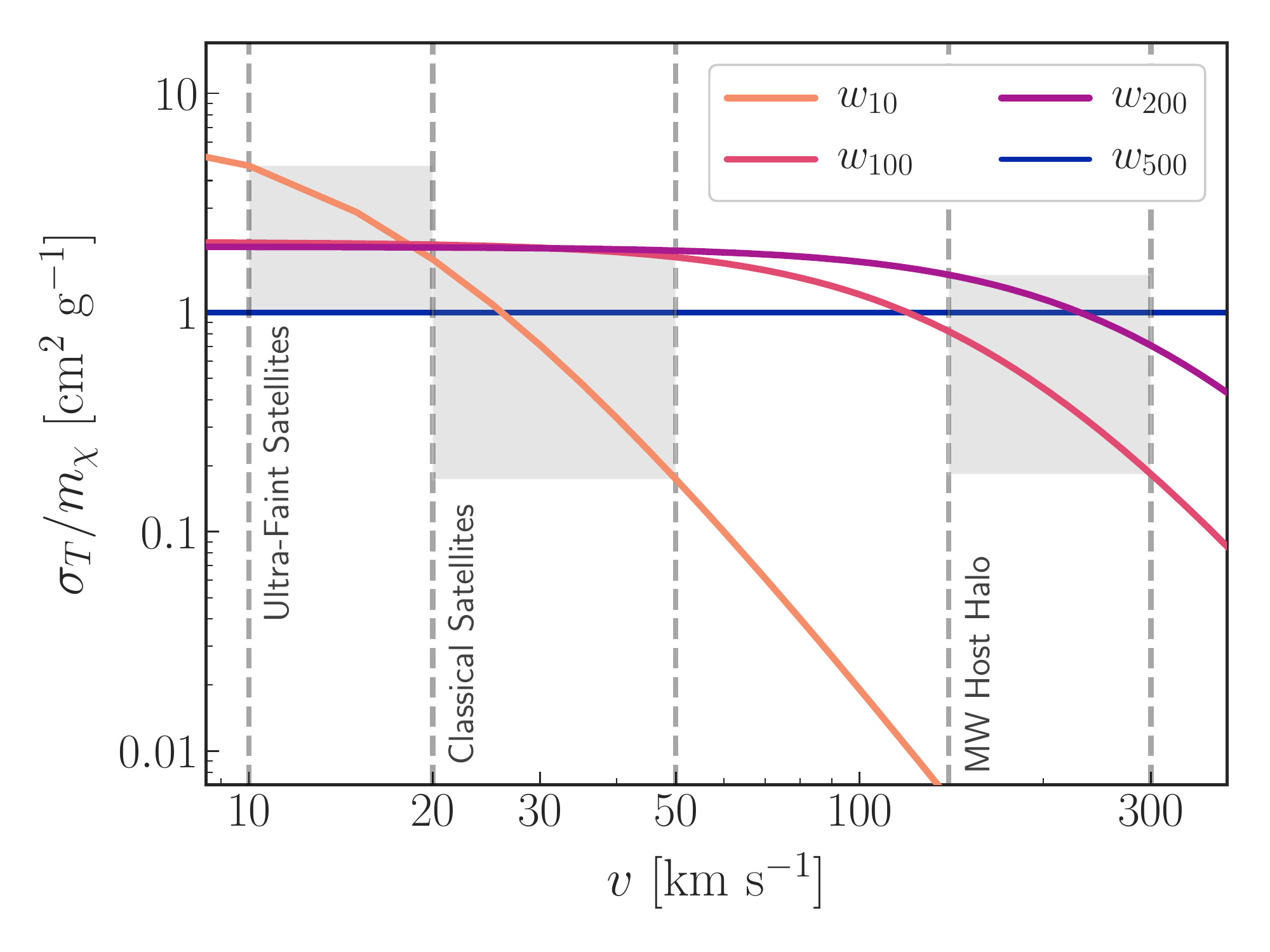}
\caption{Momentum transfer cross sections for our SIDM model variants as a function of relative scattering velocity. Each model variant is labeled by $w$, the velocity scale above which the SIDM cross section falls off as $v^{-4}$. The velocity scale relevant for interactions among host halo particles and between host halo and subhalo particles is indicated by the ``MW Host Halo'' band. Shaded bands indicate characteristic velocities for DM particles within the subhalos expected to host classical and ultrafaint MW satellite galaxies.}
\label{fig:xsec}
\end{figure}

At velocity scales $v>w$, the momentum transfer cross section falls off as $v^{-4}$; meanwhile, for $v\ll w$, it flattens toward its asymptotic value. Thus, for values of $w$ that are small relative to the typical velocities in a virialized system, interactions at low relative velocities are more effective at transferring momentum than interactions at high relative velocities. In the context of MW-mass systems, this implies that interactions among host halo particles and between host halo and subhalo particles are less effective than interactions among subhalo particles if $w$ is small relative to the characteristic velocity scale of the host~($\sim 200\ \rm{km\ s}^{-1}$). On the other hand, as $w$ increases toward the velocity scale of the host halo, subhalo--host halo and host halo--host halo interactions become more significant.

The characteristic velocity scale $w$ is an easily interpretable quantity that captures much of the key physics that shapes MW-mass systems in our two-parameter SIDM model. Thus, we parameterize and refer to our simulations in terms of $w$, and we choose $\sigma_0/m_\chi$ for each model so that the momentum transfer cross section at the velocity scales of interest yields enough self-interactions to produce observable effects, but not so many that the models are likely ruled out already. In particular, we study the four following SIDM model variants:
\begin{enumerate}
\item $w_{10}$: a model with $w=10\ \rm{km\ s}^{-1}$, $\sigma_0/m_\chi = 8/\pi\ \rm{cm^2\ g}^{-1}$, for which self-interactions within subhalos are significant, but self-interactions at the host halo velocity scale are negligible;
\item $w_{100}$: a model with $w=100\ \rm{km\ s}^{-1}$, $\sigma_0/m_\chi = 2/\pi\ \rm{cm^2\ g}^{-1}$, for which self-interactions within low-mass subhalos are less significant than in $w_{10}$, but self-interactions at the host halo velocity scale are more significant;
\item $w_{200}$: a model with $w=200\ \rm{km\ s}^{-1}$, $\sigma_0/m_\chi = 2/\pi\ \rm{cm^2\ g}^{-1}$, for which self-interactions within low-mass subhalos are identical to $w_{100}$, but self-interactions at the host halo velocity scale are more significant;
\item $w_{500}$: a model with $w=500\ \rm{km\ s}^{-1}$, $\sigma_0/m_\chi = 1/\pi\ \rm{cm^2\ g}^{-1}$, for which self-interactions within low-mass subhalos are the least significant among our model variants, while interactions at the host halo velocity scale are similar to $w_{200}$ (though slightly more effective at high velocities). Scattering in this model is isotropic and velocity independent on the scales relevant for MW-mass systems, meaning that self-interactions with large scattering angles at high relative velocities are potentially significant.
\end{enumerate}
Figure \ref{fig:xsec} shows the momentum transfer cross section corresponding to each model variant, and their main properties are summarized in Table \ref{tab:sidm_sims}.

\section{Simulations}
\label{sec:simulations}

\subsection{General Description}

For each SIDM model variant, we resimulate the same MW-mass halo using fixed initial conditions. This host halo is chosen from the suite of CDM-only MW-mass zoom-in simulations presented in \citet{Mao150302637}. The highest-resolution particles in our zoom-in simulations have a mass of $3\times 10^{5}\ M_{\rm \odot}\ h^{-1}$, and the softening length in the highest-resolution regions is $170\ \text{pc}\ h^{-1}$. Halo catalogs and merger trees were generated using the {\sc Rockstar} halo finder and the {\sc consistent-trees} merger tree code \citep{Behroozi11104372,Behroozi11104370}. Throughout we define virial quantities using the \cite{Bryan9710107} critical overdensity~$\Delta_{\rm vir}\simeq 99.2$, as appropriate for the cosmological parameters in our simulations: $h = 0.7$,~$\Omega_{\rm m} = 0.286$, $\Omega_{\rm b} = 0.047$, and $\Omega_{\Lambda} = 0.714$.

\begin{deluxetable*}{{l@{\hspace{0.08in}}c@{\hspace{0.08in}}c@{\hspace{0.08in}}c@{\hspace{0.08in}}c@{\hspace{0.08in}}c@{\hspace{0.08in}}c}}[t]
\centering
\tablecolumns{7}
\tablecaption{Summary of SIDM Model Variants and Simulation Results.}
\tablehead{
\colhead{Simulation\phantom{textttttt}} & \colhead{$w\ (\rm{km\ s}^{-1})$\phantom{tex}} & \colhead{$\sigma_0/m_\chi\ (\rm{cm^2\ g}^{-1})$} & \colhead{Cored Host Halo?} & \colhead{Cored Subhalos?} & \colhead{Ram Pressure Stripping?} & \colhead{$N_{\rm{SIDM}}/N_{\rm{CDM}}$} }
\startdata
CDM
& $\cdots$ 
& $\cdots$
& \text{\sffamily X} 
& \text{\sffamily X}
& \text{\sffamily X}
& $1.0$\\
$w_{10}$
& $10$ 
& $8/\pi$ 
& \text{\sffamily X} 
& \checkmark
& \text{\sffamily X}
& $0.82$\\
$w_{100}$
& $100$ 
& $2/\pi$ 
& \checkmark 
& \checkmark
& \checkmark
& $0.64$\\
$w_{200}$
& $200$ 
& $2/\pi$ 
& \checkmark 
& \checkmark
& \checkmark
& $0.65$\\
$w_{500}$
& $500$ 
& $1/\pi$ 
& \checkmark
& \checkmark
& \checkmark
& $0.44$\\
\enddata
{\footnotesize \tablecomments{SIDM model variants considered in this work and the main qualitative results of our zoom-in simulation of an MW-mass system for each case. The first column lists the DM model, the second and third columns list the characteristic velocity scale and amplitude of the self-interaction cross section for our SIDM model variants, the fourth (fifth) column lists whether the host halo (subhalos) are cored by self-interactions, the sixth column lists whether subhalos are affected by ram pressure stripping due to self-interactions with the host, and the seventh column lists the fraction of subhalos among our matched subhalo populations that survive to $z=0$ relative to the number of surviving subhalos with $V_{\rm{peak}}>20\ \rm{km\ s}^{-1}$ in our CDM simulation.}}
\label{tab:sidm_sims}
\end{deluxetable*}

\subsection{SIDM Implementation}
\label{sec:implementation}

To implement DM self-interactions in our zoom-in simulations, we follow the prescription in \cite{Banerjee190612026}. Briefly, self-interactions are implemented using a modified version of the {\sc GADGET-2} $N$-body code that allows DM particles to transfer momentum and energy with an interaction probability set by the total SIDM cross section in Equation \ref{eq:sigma_tot} and a scattering angle drawn from the distribution given by Equation \ref{eq:xsec}. We refer the reader to \cite{Banerjee190612026} for the details of our SIDM implementation.

\subsection{Subhalo Definitions and Resolution Cuts}
\label{sec:defs}

Throughout we employ the following definitions when referring to ``subhalos'':
\begin{enumerate}
\item Surviving subhalos: DM systems identified by {\sc Rockstar} as distinct bound objects within the virial radius of the MW host halo at $z=0$;
\item Disrupted subhalos: DM systems that have crossed within the virial radius of the MW halo at any simulation snapshot (where the virial radius of the MW halo is evaluated as a function of time) but are no longer identified as distinct bound objects by {\sc Rockstar} at $z=0$ because they have deposited the majority of their particles onto the main-branch progenitor of the host at any earlier snapshot. Operationally, we require the descendant of each disrupted subhalo to be on the main branch of the host, following \cite{Nadler171204467}.\footnote{For concreteness, we find that the average virial mass of disrupted subhalos in our CDM simulation is $\sim 10^8\ M_{\mathrm{\odot}}$ at the time of disruption, and that $\sim 90\%$ of these disrupting subhalos have virial masses below $10^9\ M_{\mathrm{\odot}}$. These values do not change significantly in our SIDM simulations.}
\end{enumerate}

Subhalos in our CDM simulation are well resolved down to a maximum peak circular velocity of $V_{\rm peak} \approx 10\ \rm{km\ s}^{-1}$, where $V_{\rm{peak}}$ is the largest maximum circular velocity a halo attains over its entire history (\citealt{Mao150302637,Nadler180905542}). However, we caution that {\sc Rockstar} is optimized for halo finding in collisionless DM simulations and therefore might be unable to reliably identify the more diffuse halos present in SIDM simulations, particularly near the resolution limit (X.\ Du, A.\ Peter, \& C.\ Zeng 2019, private communication). Thus, we study subhalos above a very conservative resolution limit of $V_{\rm peak} > 20\ \rm{km\ s}^{-1}$ in our CDM simulation. This cut is applied to both surviving and disrupted subhalos, and it corresponds to subhalos resolved with more than~$\sim 1000$ particles at the time $V_{\mathrm{peak}}$ is achieved.

Due to this conservative resolution cut, we expect that artificial subhalo disruption (e.g., \citealt{VandenBosch171105276,VandenBosch180105427}) is not a large source of error relative to accurate halo finding in the analysis of our SIDM simulations. It is important to note that our working definition of ``subhalo disruption'' cannot distinguish truly unbound systems from those that fall below the resolution limit of our simulations or the detection capabilities of our halo finder. Nonetheless, convergence tests based on higher-resolution resimulations of CDM and $w_{500}$ presented in Appendix \ref{appendixa} confirm that the trends reported in this paper, and particularly the correlation between $w$ and the severity of subhalo disruption, are robust to the resolution of our simulations.

\begin{figure*}[t]
\centering
\includegraphics[scale=0.52]{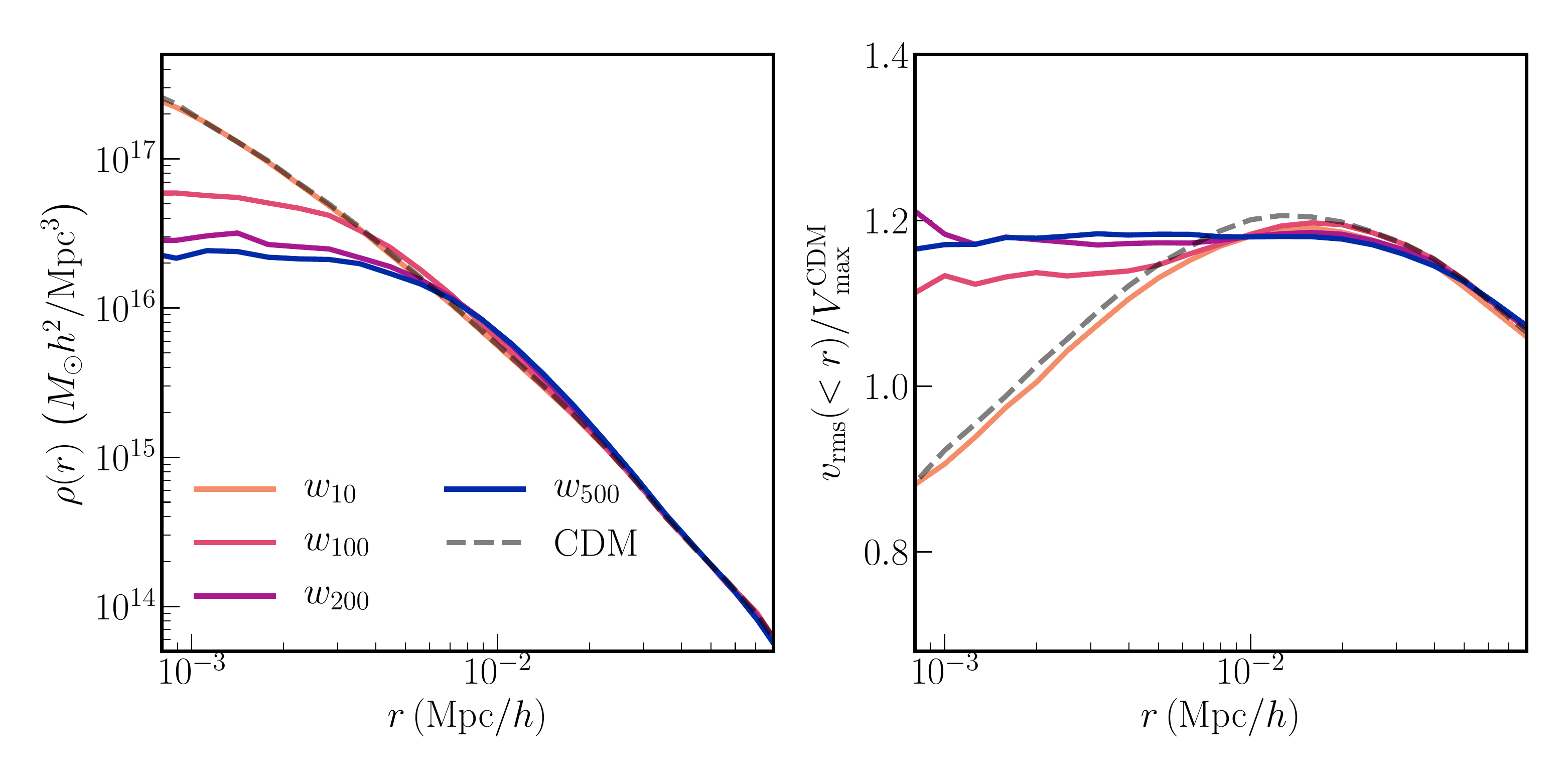}
\caption{SIDM effects on the host halo. Left panel: host halo density profiles for our SIDM model variants. Right panel: corresponding velocity dispersion profiles. As the SIDM cross section at the characteristic velocity scale of the MW host halo increases, the inner regions of the host become increasingly cored and thermalized. Note that the host halo density and velocity dispersion profiles are nearly indistinguishable in CDM and $w_{10}$.}
\label{fig:host_properties}
\end{figure*}

\subsection{Subhalo Matching Procedure}
\label{sec:matching}

For a fixed $V_{\rm{peak}}$ threshold, different numbers of surviving and disrupted subhalos may exist in our CDM and SIDM simulations at $z=0$. Thus, to ensure that we analyze the same population of subhalos in CDM and SIDM, we adopt a variable $V_{\rm{peak}}$ threshold for our SIDM simulations, denoted by $V_{\rm{thresh}}$. We define $V_{\rm{thresh}}$ such that the number of surviving-plus-disrupted subhalos with~$V_{\rm{peak}}>V_{\rm{thresh}}$ is equal to the number of such subhalos in CDM with~$V_{\rm peak}>20\ \rm{km\ s}^{-1}$.\footnote{There are $154$ surviving and disrupted subhalos in our CDM simulation with $V_{\rm{peak}}>20\ \rm{km\ s}^{-1}$.} This yields thresholds of $V_{\rm{thresh}}=[19.5,19.05,18.75,18.7]\ \rm{km\ s}^{-1}$ for $[w_{10},w_{100},w_{200},w_{500}]$, respectively. We have verified that these resolution cuts result in the same population of subhalos on an object-by-object basis by inspecting the initial phase-space region corresponding to each subhalo. We explore less conservative subhalo resolution thresholds in Appendix~\ref{appendixb}, where we show that our main findings are robust to the chosen $V_{\rm{peak}}$ threshold.

We reiterate that authors have used various halo-finding algorithms and SIDM implementations to study the effects of self-interactions in MW-mass systems. Thus, direct comparisons of different simulations should be performed with care, particularly near the resolution limit, where halo finder shortcomings and other spurious numerical effects are expected to be most severe. We compare our findings to previous results in Section \ref{subsec:comparison}.

\section{SIDM Effects on the MW Host Halo}
\label{subsec:host}

We now present our results, focusing on the properties of the MW host halo and its subhalo population in each SIDM model variant described above. Table \ref{tab:sidm_sims} lists the main qualitative results of each simulation.

The virial mass of the MW host halo is virtually identical in all of our simulations. However, the DM profile in the inner regions of the host varies significantly as a function of~$w$. This is illustrated in Figure~\ref{fig:host_properties}, which shows the density and velocity dispersion profiles of the host in CDM and in each of our SIDM model variants. As expected, SIDM models with larger self-interaction cross sections at the velocity scale set by the host's velocity dispersion ($\sim 200\ \rm{km\ s}^{-1}$) exhibit cored density profiles, and the DM distribution in the inner regions of these hosts is roughly isothermal, consistent with previous findings (e.g., \citealt{Dave0006218,Colin0205322,Vogelsberger12015892,Vogelsberger180503203,Rocha12083025,Robles190301469}). In contrast, the host halo in $w_{10}$ is very similar to that in CDM because self-interactions at the host's velocity scale are negligible in this case.\footnote{We have also examined the ellipticity profile of the host halo in each simulation. We find that hosts in model variants with larger values of $w$ are more spherical, with typical ellipticities near unity, while the ellipticity profiles in $w_{10}$ and CDM rise from $\sim 0.9$ in the inner regions to $\sim 1$ at the virial radius; these findings are consistent with many previous studies (see, e.g., \citealt{Tulin170502358}).}

\begin{figure*}[t]
\hspace{-15mm}
\includegraphics[scale=0.72]{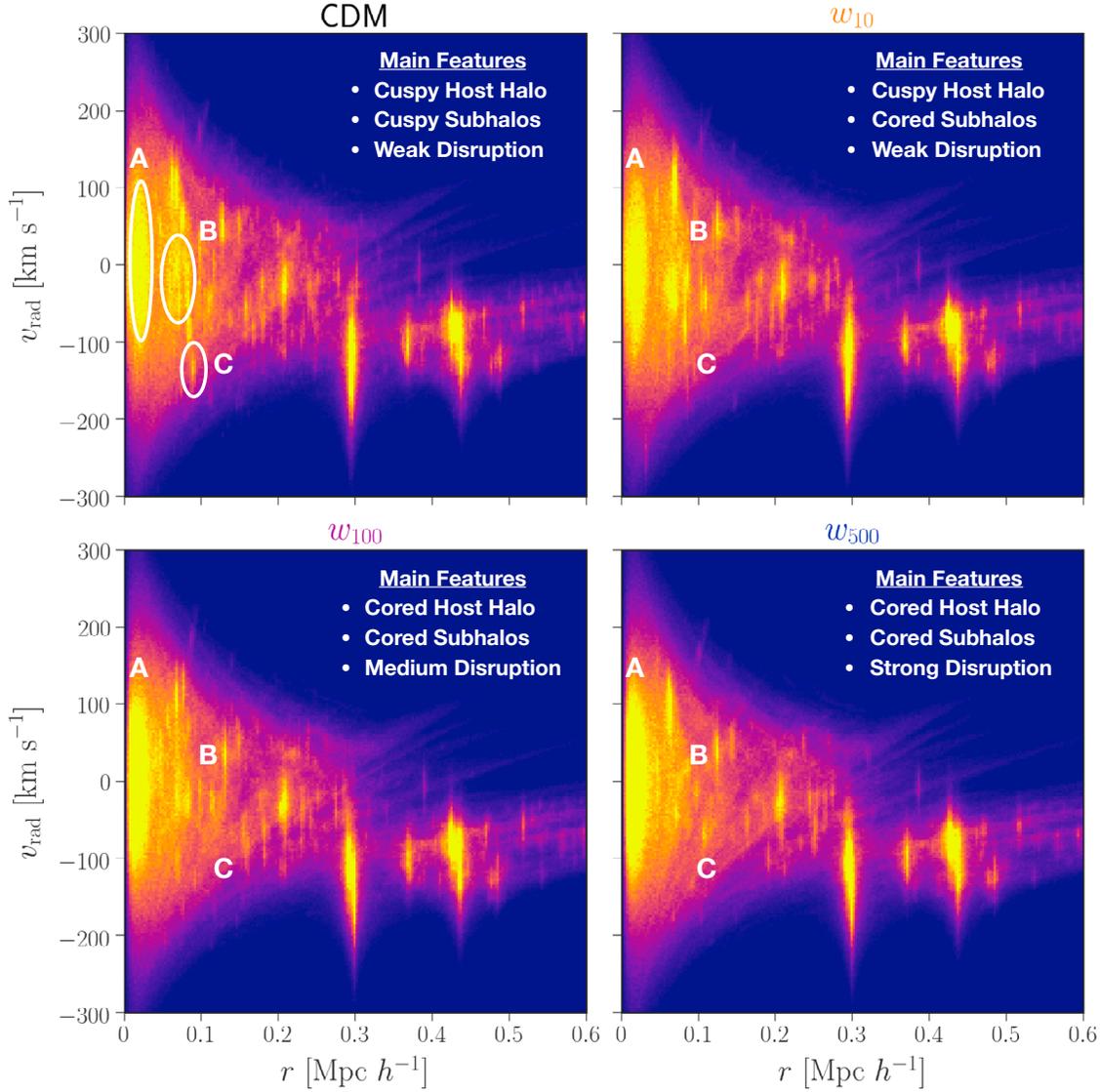}
\caption{DM phase-space distributions for our zoom-in simulations of an MW-mass host halo in CDM and SIDM. The density of DM particles in bins of radial velocity and radial distance from the center of the host is plotted for CDM (top left) and for three of our SIDM model variants:~$w_{10}$ (top right), $w_{100}$ (bottom left), and $w_{500}$ (bottom right). These distributions qualitatively illustrate several of our main findings. For example, the host halo (labeled A) has a very similar phase space distribution in CDM and $w_{10}$, while subhalos in $w_{10}$ (e.g., subhalo B) are somewhat less dense because of the large self-interaction cross section at low relative velocities in this case (see Figure \ref{fig:xsec}). On the other hand, particles near the center of the host in $w_{200}$ and $w_{500}$ are preferentially scattered onto tangential orbits, and many of the low-mass subhalos that survive in CDM (e.g., subhalo C) are disrupted in these SIDM model variants owing to a combination of ram pressure stripping caused by self-interactions with the host and tidal stripping.}
\label{fig:phase}
\end{figure*}

\begin{figure*}[t]
\includegraphics[scale=0.43]{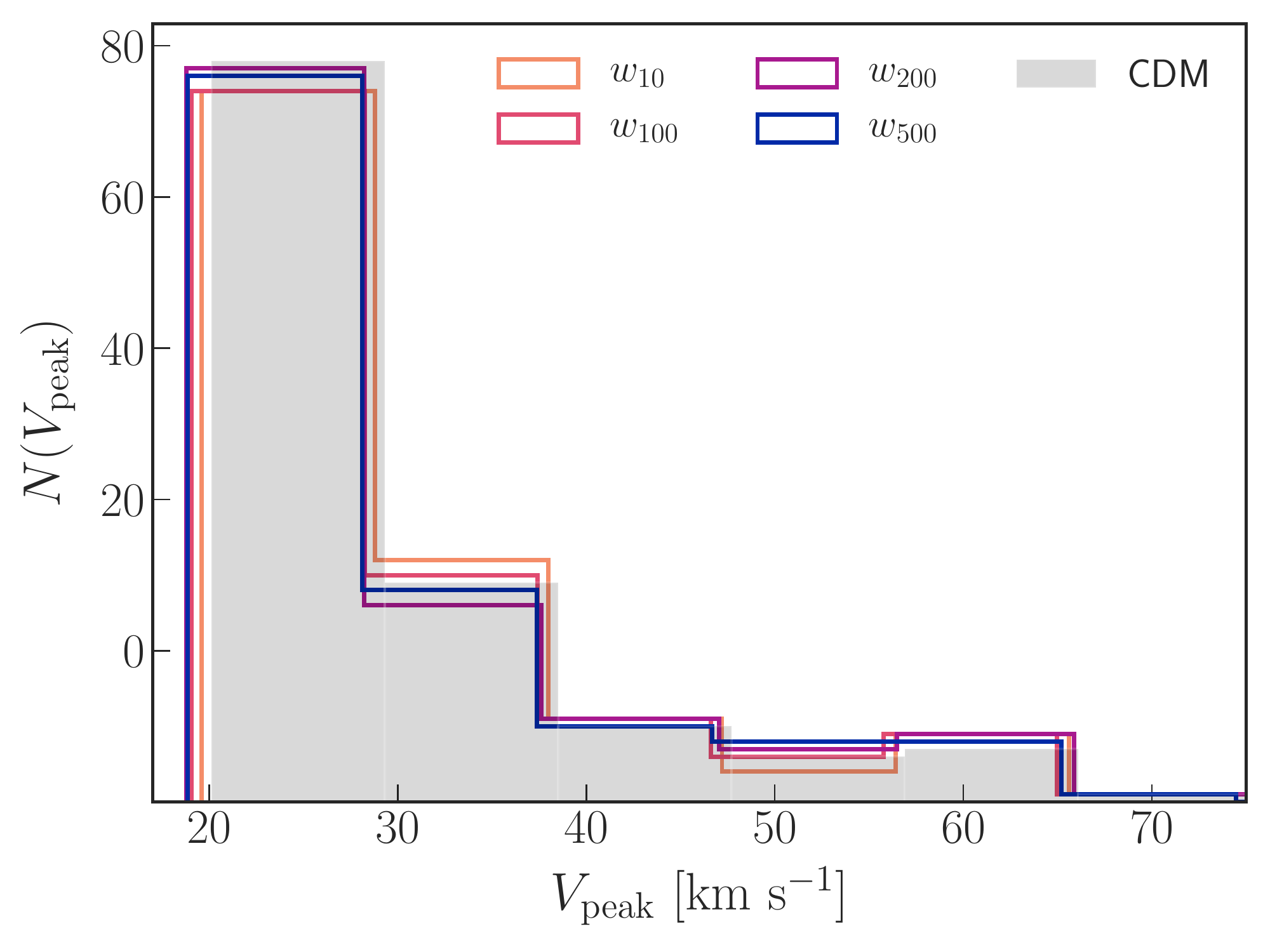}
\hspace{1.5mm}
\includegraphics[scale=0.43]{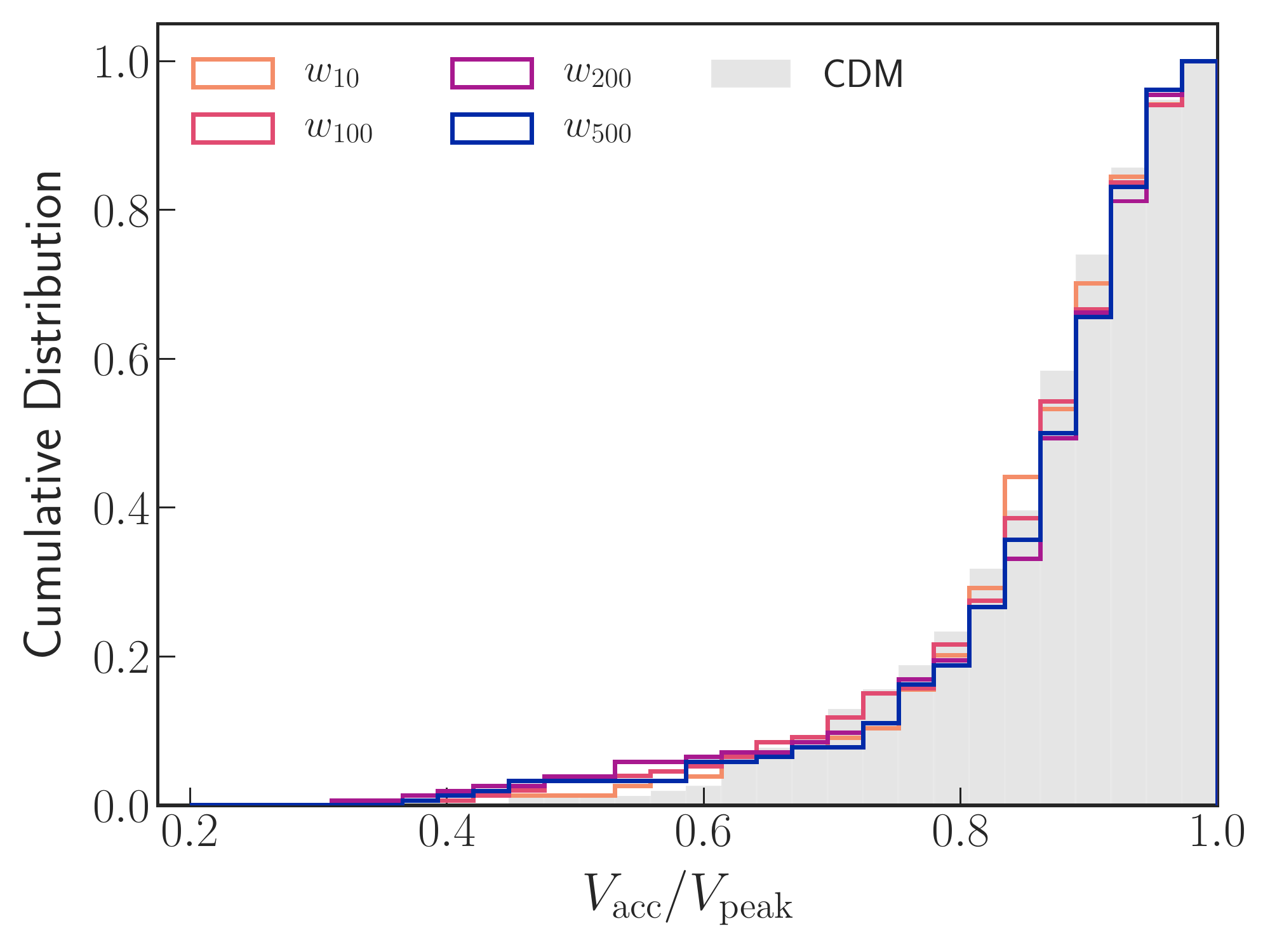}
\caption{SIDM effects on subhalos before infall. Left panel: distributions of peak maximum circular velocity for surviving and disrupted subhalos in each SIDM model variant (unfilled histograms) vs.\ CDM (filled histogram). Right panel: cumulative distributions of the ratio of maximum circular velocity evaluated at the time of each subhalo's accretion onto the host divided by the peak maximum circular velocity along the main branch of the subhalo. Although subhalo assembly is statistically identical in CDM and SIDM, subhalos are mildly stripped by self-interactions prior to accretion onto the host halo in our SIDM simulations.}
\label{fig:vpeak_hist}
\end{figure*}

As demonstrated by several authors (e.g., \citealt{Kaplinghat13116524,Sameie180109682,Robles190301469}), these results are expected to change in the presence of a central baryonic component such as the Galactic disk. In particular, because DM dynamically responds to the \emph{total} (i.e., DM-plus-baryonic) gravitational potential, we expect our host halos to exhibit much smaller cores or even cusps in the presence of baryons. Importantly, this implies that the findings discussed below on the disruption of SIDM subhalos are strictly lower limits, since tidal disruption and ram pressure stripping would be more severe for the denser, hotter host halo expected in the presence of baryons.

To visualize the present-day DM structure in our simulations, Figure~\ref{fig:phase} shows the DM particle density in the phase space of radial distance versus radial velocity with respect to the center of the host. We observe that the DM profiles of both the host halo and its subhalos are less concentrated in our SIDM simulations. This effect is more significant for larger values of~$w$; for example, visual inspection of Figure~\ref{fig:phase} suggests that many prominent substructures that survive in CDM are completely disrupted in $w_{500}$. We also find that the phase-space density at small radii and low radial velocities with respect to the host center increases with $w$, implying that particles on radial orbits (i.e., those with high radial velocities in our CDM simulation) are scattered onto tangential orbits owing to self-interactions. Finally, we note that specific substructures in $w_{10}$ appear more diffuse than their counterparts in CDM, which arises as a result of the large momentum transfer cross section at low relative velocities for $w_{10}$ (see Figure~\ref{fig:xsec}).

The differences in the host halo's density and velocity dispersion profiles discussed above impact the post-infall evolution of subhalos. In particular, it is expected that the present-day ($z=0$) abundance and properties of subhalos are affected by ram-pressure-like stripping due to self-interactions between subhalo and host halo particles, as well as by tidal effects in the host halo's gravitational field. We investigate these effects in Section \ref{sec:post-infall}.

Thus, the host halo is cored and thermalized in the absence of baryons if self-interactions are significant at the velocity scale set by the host's velocity dispersion. These effects will be weakened in the presence of baryons.

\section{SIDM Effects on MW Subhalos}
\label{subsec:sub}

Next, we examine the pre- and post-infall evolution of all subhalos that fall into the host in each of our simulations. We then compare the corresponding \emph{surviving} subhalo populations at $z=0$.

\subsection{Pre-infall Subhalo Evolution}
\label{sec:pre-infall}

In this subsection, we explore the formation and evolution of subhalos before they fall into the host.

\subsubsection{Subhalo Assembly}
\label{sec:assembly}

First, we investigate the formation and initial properties of subhalos in each of our simulations. To do so, we plot the distribution of $V_{\rm{peak}}$ for each SIDM model variant in the left panel of Figure~\ref{fig:vpeak_hist}, and we compare these to CDM. The $V_{\rm{peak}}$ distributions are calculated using the matched subhalo populations defined in Section \ref{sec:matching}, and they include both subhalos that survive to $z=0$ and those that fall into the host halo before disrupting. 
We find that the $V_{\rm{peak}}$ distributions are unchanged relative to CDM. In addition, the distributions of the time at which $V_{\rm{peak}}$ is achieved are nearly identical among the simulations, although there is a small amount of scatter on a subhalo-by-subhalo basis.

Thus, the formation times and initial properties of subhalos in all of our SIDM simulations---defined in terms of $V_{\rm{peak}}$ and the time at which $V_{\rm{peak}}$ occurs---are statistically identical to the corresponding quantities in our CDM simulation.


\subsubsection{Effects of Early Self-interactions}
\label{sec:pre-infall-evolution}

Next, we assess the effects of self-interactions on subhalos before infall into the host. In particular, the right panel of Figure \ref{fig:vpeak_hist} shows the cumulative distribution of maximum circular velocity evaluated at the time of accretion onto the host, $V_{\rm{acc}}$, divided by $V_{\rm{peak}}$.\footnote{Accretion is defined as the snapshot at which the center of a subhalo crosses into the virial radius of the host. Note that the virial radius is measured as a function of time using the {\sc Rockstar} output.} This quantity captures the impact of self-interactions between the early time at which $V_{\rm{peak}}$ is usually achieved ($z_{\rm{peak}}\sim 3$ for surviving subhalos, on average) and the time of accretion ($ z_{\rm{acc}} \sim 1$ for surviving subhalos, on average). These characteristic peak and infall times are very similar among our CDM and SIDM simulations.

We observe a subtle but systematic trend in $V_{\rm{acc}}/V_{\rm{peak}}$ as a function of the characteristic self-interaction velocity scale~$w$. In particular, the low-$V_{\rm{acc}}/V_{\rm{peak}}$ tail of this distribution, which is present in CDM and results from pre-infall tidal stripping (e.g., \citealt{Behroozi13102239,Wetzel150101972}), is more prominent in SIDM. This difference becomes increasingly pronounced for SIDM model variants with larger values of~$w$, indicating that self-interactions at the typical relative velocity scale affect subhalos during pre-infall tidal stripping.

This finding is consistent with the fact that regions outside the conventional virial radius of the host halo can be quite dense (note that the self-interaction rate depends on the physical density, rather than the comoving density). In particular, typical splashback boundaries for MW-mass halos extend to $\sim 1.5$ times the conventional virial radius (e.g., \citealt{Adhikari14094482,Diemer:2014xya,More150405591}). Accordingly, we find that there is a noticeably smaller difference in the low-$V_{\rm{acc}}/V_{\rm{peak}}$ tail relative to CDM if $V_{\rm{acc}}$ is evaluated when subhalos cross within twice the virial radius of the host.

In Section \ref{sec:subhalo_profiles}, we show that the density profiles of subhalos in our SIDM model variants that exhibit appreciable changes to their $V_{\rm{acc}}/V_{\rm{peak}}$ distribution relative to CDM (i.e., all model variants other than $w_{10}$) already exhibit cored density profiles at infall. We provide a physical interpretation of the scaling with $w$ in Section \ref{sec:effects}.

Thus, subhalos in our SIDM simulations are affected by self-interactions before accretion onto the host, leading to lower values of $V_{\rm{acc}}/V_{\rm{peak}}$ relative to subhalos in CDM.

\subsection{Post-infall Subhalo Evolution}
\label{sec:post-infall}

We now explore the evolution of subhalos in our SIDM model variants after infall into the host. In particular, we describe the physical effects that influence subhalos after infall, and we investigate subhalo disruption in our simulations.

\subsubsection{Physical Effects}
\label{sec:effects}

After infall, two main effects shape the subhalo populations in our SIDM simulations:
\begin{enumerate}
\item \emph{Tidal stripping}: The gravitational field of the host halo tidally strips material from subhalos. Note that this effect is velocity independent; in particular, assuming spherical symmetry and considering tidal interactions with the host halo only, the mass-loss rate due to tidal stripping can be written as \citep{VandenBosch171105276}
\begin{equation}
\left(\frac{\dot{M}_{\rm{sub}}}{M_{\rm{sub}}}\right)_{\rm{tidal}} = -\frac{1}{\alpha}\frac{1}{\tau_{\rm{orb}}}\frac{M_{\rm{sub}}(>r_t)}{M_{\rm{sub}}},
\end{equation}
where $M(>r_t)$ is the mass of a subhalo contained outside of its tidal radius, $\tau_{\rm{orb}}$ is the orbital timescale, and $\alpha$ is an order-unity constant. Note that the tidal radius (i.e., the distance from the center of the subhalo at which the host's tidal force is balanced by the subhalo's own gravity) depends on both the host halo and subhalo density profiles and the distance of the subhalo from the center of the host. Although tidal stripping occurs in CDM-only simulations, differences in the strength of this effect may arise in SIDM owing to changes in the density profiles of both the host halo and its subhalos caused by self-interactions. In particular, we expect that the cored density profile of the host halo in SIDM reduces the efficacy of tidal stripping relative to CDM for a fixed subhalo profile. However, as discussed in, e.g., \cite{Kahlhoefer190410539}, the cores induced in SIDM subhalos make them more susceptible to disruption in a fixed tidal field.

\item \emph{Ram pressure stripping}: Self-interactions with host halo particles drive material out of subhalos. Taken to the extreme, this process can completely dissociate subhalos in a phenomenon known as subhalo evaporation (see, e.g., \citealt{Vogelsberger180503203}). Unlike tidal stripping, which strips mass from the outskirts of subhalos, ram pressure interactions remove material throughout their extent. In particular, subhalos experience an effective pressure owing to interactions with the host; the mass-loss rate due to these ram-pressure-like interactions can be written as \citep{Kummer170604794}
\begin{equation}
    \left(\frac{\dot{M}_{\rm{sub}}}{M_{\rm{sub}}}\right)_{\rm{ram-pressure}} = -\chi_e(v_{\rm{sub}},v_{\rm{esc,sub}})\rho_{\rm{host}}(\mathbf{r}) v_{\rm{sub}}(\mathbf{r})\frac{\sigma(v_{\rm{sub}})}{m_\chi},\label{eq:ram-pressure}
\end{equation}
where $\chi_e$ is an order-unity factor, $v_{\rm{sub}}(\mathbf{r})$ is the velocity of the subhalo relative to the host evaluated along its orbit, $v_{\rm{esc,sub}}$ is the escape velocity from the subhalo (which depends on the subhalo's density profile), $\rho_{\rm{host}}(\mathbf{r})$ is the density of the host halo evaluated along the subhalo's orbit, and $\sigma(v_{\rm{sub}})/m_\chi$ is the total SIDM cross section evaluated at $v_{\rm{sub}}$.

\hspace{0.08in} The mass-loss rate due to ram pressure stripping depends on the SIDM cross section at the typical relative velocity scale between the host halo and its subhalos, which is set by the gravitational potential of the host. Because SIDM models with larger values of $w$ have larger self-interaction cross sections at this velocity scale, the strength of ram pressure stripping increases with $w$.
\end{enumerate}

By integrating Equation \ref{eq:ram-pressure} over the course of a typical orbit, we find that subhalos with close pericentric passages to the center of the host ($d_{\rm{peri}}\lesssim 70\ \rm{kpc}$) can lose $\sim 10\%$ of their infall mass owing to ram pressure stripping alone, for an isotropic cross section of $\sigma/m_\chi=2~ \rm{cm^2\ g}^{-1}$. Combined with tidal effects, and particularly the fact that subhalos with cored density profiles are more susceptible to tidal disruption, our calculations suggest that self-interactions at the relevant velocity scales are sufficient to severely strip subhalos in SIDM model variants with large values of $w$. We have confirmed that this behavior---i.e., increased mass loss relative to CDM followed by disruption due to tidal forces at pericenter---is indeed the dominant disruption mechanism in our simulations by inspecting the histories of matched subhalos that survive in CDM but disrupt in our SIDM model variants.

Thus, the evolution of subhalos after infall in our SIDM simulations is driven by a combination of tidal forces and ram-pressure stripping. Ram-pressure stripping is more effective for models with larger values of $w$, which have larger self-interaction cross sections at the host halo velocity scale. This mass-loss mechanism makes subhalos more susceptible to tidal disruption, particularly during pericentric passages.

\subsubsection{Subhalo Disruption}
\label{sec:disruption}

We now analyze our simulation results to quantitatively assess the impact of the effects described in the previous section. In our CDM simulation, roughly half of all subhalos with $V_{\rm{peak}}>20\ \rm{km\ s}^{-1}$ that fell into the host disrupt by $z=0$. We find that even larger fractions of subhalos disrupt in our SIDM simulations, as indicated by the surviving subhalo fractions in Table \ref{tab:sidm_sims}. Based on the arguments in Section \ref{sec:effects}, we expect that extra subhalo disruption relative to CDM should be correlated with the SIDM cross section at the typical relative velocity scale set by the host's gravitational potential, such that subhalos are disrupted more efficiently in SIDM model variants with larger values of $w$. As demonstrated below, our results are consistent with this hypothesis.

\begin{figure*}[t]
\centering
\includegraphics[scale=0.3]{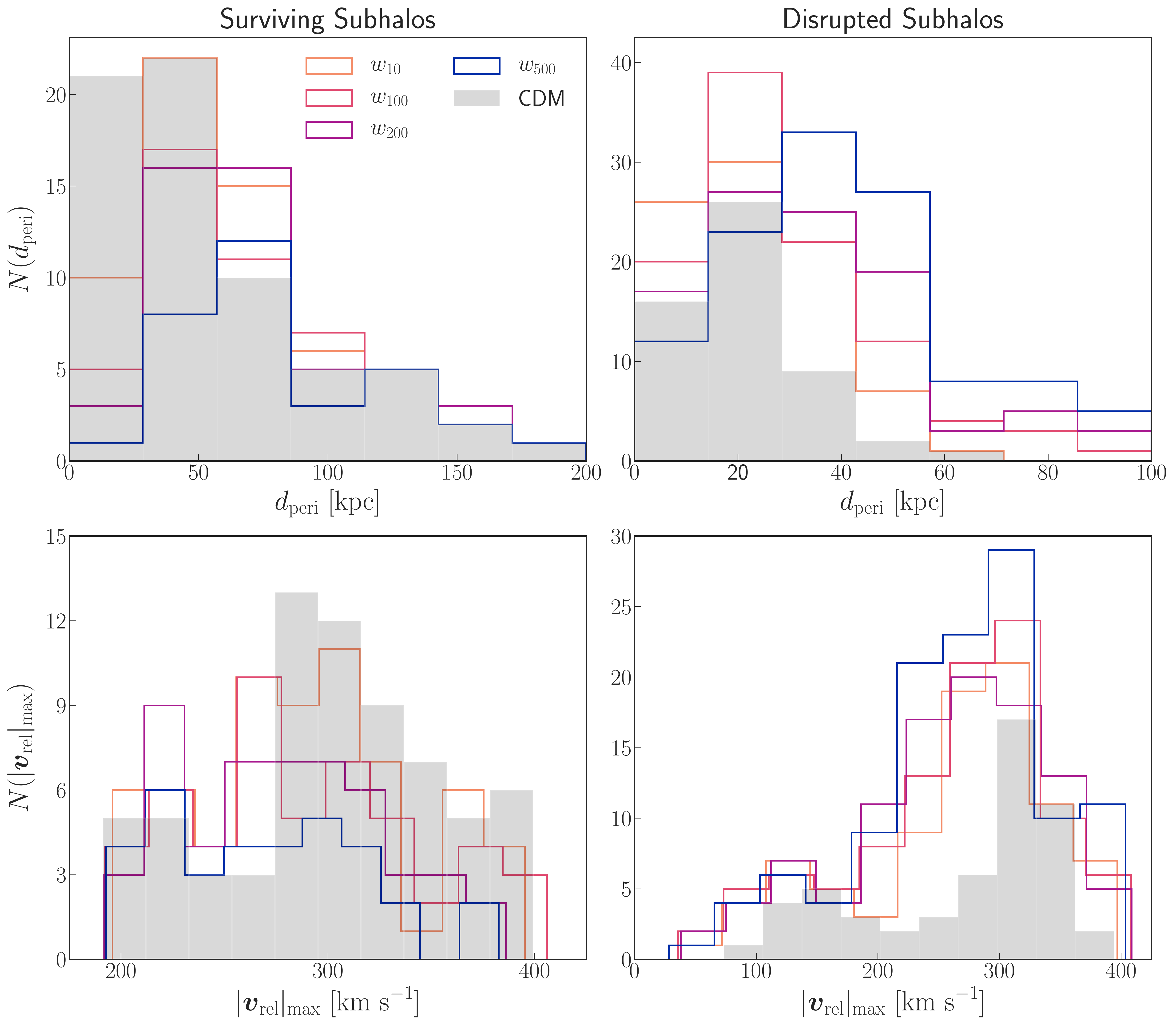}
\caption{Properties of surviving and disrupted subhalos. Top left panel: distribution of the distance of closest approach to the host at first pericentric passage for surviving subhalos in each of our SIDM model variants (open histograms) vs.\ CDM (filled histogram). Top right panel: same as the top left panel, but for disrupted subhalos (surviving and disrupted subhalos are defined in Section \ref{sec:defs}). Bottom left panel: distribution of the maximum relative velocity with respect to the host halo evaluated along the orbit of each surviving subhalo. Bottom right panel: same as the bottom left panel, but for disrupted subhalos. Many subhalos that survive in our CDM simulation disrupt during early pericentric passages in our SIDM simulations, because ram pressure stripping caused by self-interactions with the host at large relative velocities makes subhalos more susceptible to tidal disruption.}
\label{fig:dperi_hist}
\end{figure*}

To build intuition about subhalo disruption in our SIDM model, we first investigate the distribution of pericentric distance, $d_{\rm{peri}}$, for disrupted subhalos in each simulation. We define $d_{\rm{peri}}$ as the distance of closest approach to the center of the host halo during each subhalo's first orbit around the host after infall.\footnote{For infalling subhalos that have not completed a pericentric passage, we set $d_{\rm{peri}}$ equal to the distance from the center of the host at $z=0$ (for surviving subhalos) or at the time of disruption (for disrupted subhalos).} The top left and top right panels of Figure~\ref{fig:dperi_hist} show the distributions of~$d_{\rm{peri}}$ for surviving and disrupted subhalos in each of our simulations. We observe that the amount of subhalo disruption at \emph{large} pericentric distances increases strongly as a function of~$w$, and we have verified that most of this disruption occurs during early pericentric passages. Subhalos in our SIDM simulations therefore disrupt at distances at which subhalos in our CDM simulation almost \emph{never} disrupt, indicating that disruption is driven by tidal effects during early pericentric passages that are enhanced by ram-pressure-like interactions with the host.

To explore the data further, the bottom left and bottom right panels of Figure \ref{fig:dperi_hist} show the distributions of maximum relative velocity with respect to the host evaluated along the orbit of each surviving and disrupted subhalo. The maximum relative velocity distributions for disrupted subhalos in our SIDM model variants peak strongly relative to CDM above~$\sim 200\ \mathrm{km\ s}^{-1}$, and the strength of this effect is correlated with $w$. This demonstrates that extra subhalo disruption relative to CDM is mainly driven by self-interactions at relative velocities above $\sim 200\ \mathrm{km\ s}^{-1}$, and that subhalos are more easily disrupted in SIDM model variants with larger momentum transfer cross sections at these velocity scales.\footnote{The bottom left panel of Figure \ref{fig:dperi_hist} also indicates that a small fraction of surviving subhalos are impacted by ram pressure stripping owing to high-velocity interactions; we return to this point in Section \ref{sec:subhalo_profiles}.}

\begin{figure*}[t]
\includegraphics[scale=0.35]{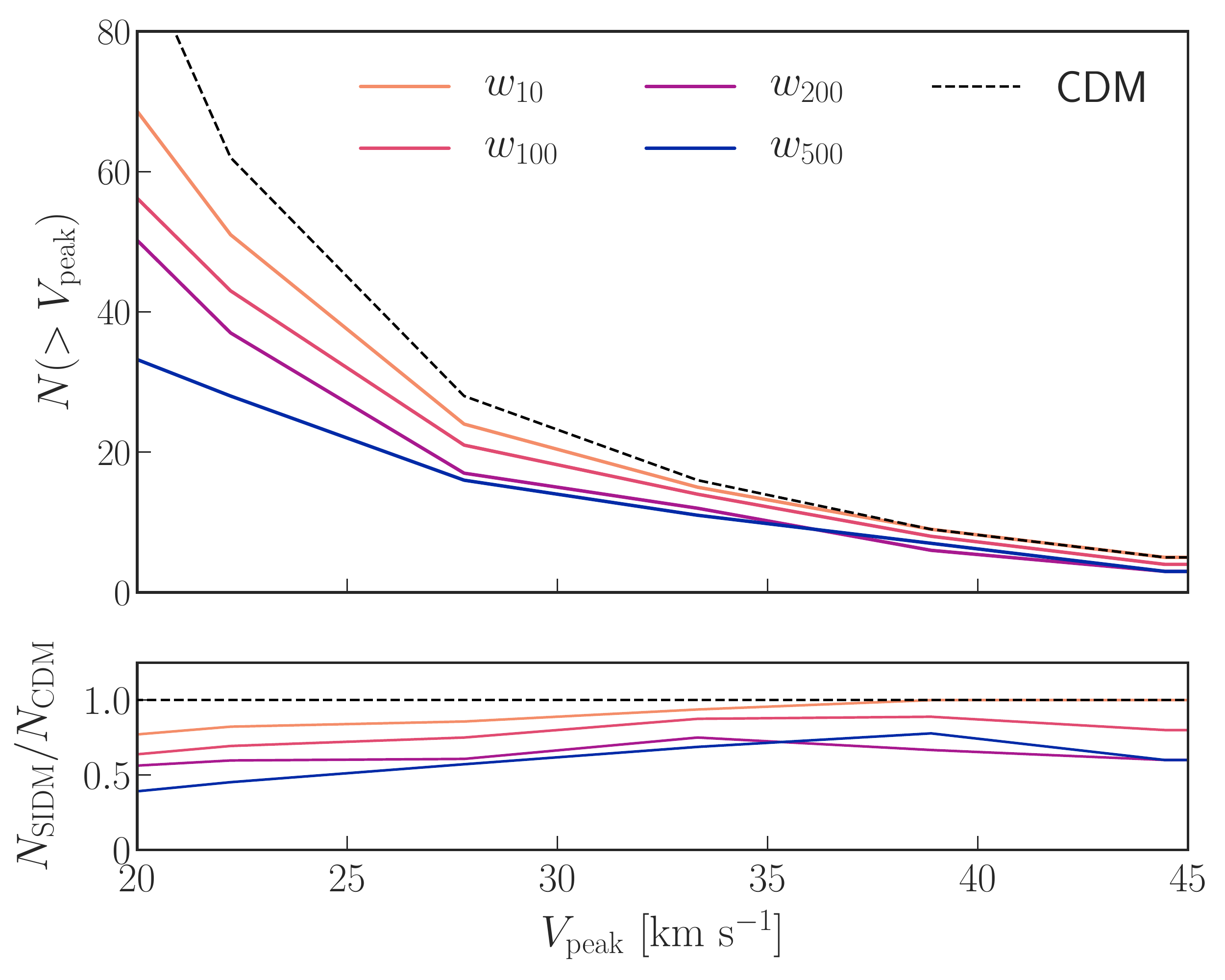}
\includegraphics[scale=0.35]{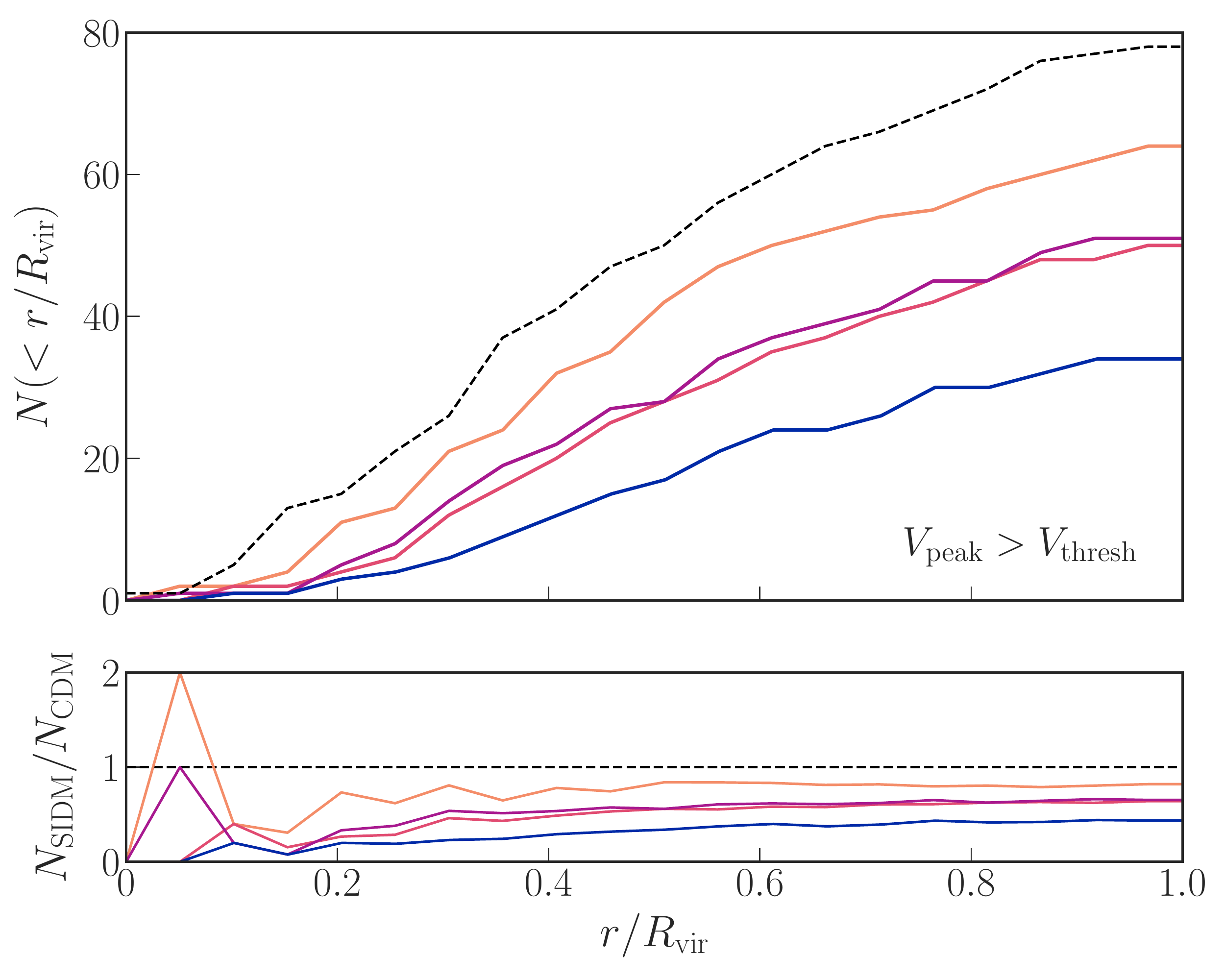}
\caption{Surviving subhalo populations. Left panel: peak velocity function of subhalos in our CDM simulation and in each of our SIDM model variants. Right panel: corresponding radial subhalo distributions in units of the host halo virial radius in each simulation. The abundance of surviving subhalos is reduced in SIDM, and the strength of this effect increases with $w$ owing to more significant ram pressure stripping caused by self-interactions with the host. Subhalo disruption in our SIDM simulations is particularly severe in the inner regions of the host halo.}
\label{fig:vsidm}
\end{figure*}

This picture is consistent with the fact that $w_{500}$---which has the largest momentum transfer cross section at large relative velocity scales---exhibits the most subhalo disruption among our SIDM model variants (see Section~\ref{sec:population_stats}). Thus, our findings support the hypothesis that a combination of ram pressure and tidal stripping increases the amount of subhalo disruption relative to CDM.

Thus, a significant fraction of the subhalos that survive in our CDM simulation are disrupted in our SIDM simulations due to extra mass loss caused by ram-pressure stripping from self-interactions with the host at large relative velocities, which makes these subhalos more susceptible to tidal disruption. This effect is more severe for models with larger values of $w$.

\subsection{Surviving Subhalo Populations}

We now examine the population statistics and properties of surviving subhalos in our SIDM simulations.

\subsubsection{Population Statistics}
\label{sec:population_stats}

In the left panel of Figure \ref{fig:vsidm}, we plot the cumulative number of surviving subhalos as a function of $V_{\rm{peak}}$ in each of our simulations. We find that the abundance of surviving subhalos monotonically decreases as a function of $w$. In particular, subhalo abundances are re-scaled in an approximately $V_{\rm{peak}}$-independent fashion, and the number of surviving subhalos in SIDM divided by the number of surviving subhalos in CDM ranges from $\sim 0.8$ in $w_{10}$ to $\sim 0.4$ in $w_{500}$. We have chosen to present these results in terms of $V_{\rm{peak}}$ because this quantity is expected to correlate more directly with satellite luminosity than, e.g., present-day virial mass or maximum circular velocity (e.g., \citealt{Reddick12072160,Lehmann151005651,Nadler180905542}, \citealt*{PaperII}).

We have verified that similar trends hold for the present-day $M_{\rm{vir}}$ and $V_{\rm{max}}$ functions; however, because these distributions mix the pre-infall properties and post-infall evolution of surviving subhalos, their shapes differ in detail from the corresponding peak functions. For example, we observe an \emph{enhancement} in subhalo abundance at large values of $V_{\rm{max}}$ for all of our SIDM model variants relative to CDM. We speculate that this is a consequence of less effective tidal stripping for surviving subhalos in our SIDM simulations, which occupy tangential orbits around a cored host halo. Previous authors have also found hints of this trend (e.g., \citealt{Rocha12083025}; also see Section \ref{subsec:comparison}).

In the right panel of Figure \ref{fig:vsidm}, we plot the radial distribution of surviving subhalos in each simulation. Similar to the~$V_{\rm{peak}}$ functions, we find that the radial distributions are approximately rescaled in the outer regions. However, disruption becomes increasingly severe near the center of the host (i.e.,~$r/R_{\rm{vir}}\lesssim 0.5$), causing $N_{\rm{SIDM}}/N_{\rm{CDM}}$ to decrease sharply in that region, except for the ``spikes'' at small radii observed for $w_{10}$ and $w_{200}$. These correspond to either one or two additional subhalos near the center of these hosts; we investigate these subhalos further in Appendix \ref{appendixc}.

Thus, subhalo disruption due to self-interactions approximately re-scales the number of surviving subhalos relative to CDM as a function of $V_{\rm{peak}}$. Subhalo disruption is particularly effective in the inner regions of the host halo, and it is more severe for models with larger values of $w$.

\subsubsection{Orbital Anisotropy Profile}
An immediate consequence of the ram-pressure-plus-tidal-stripping disruption mechanism described above is that subhalos on radial orbits are preferentially disrupted. We therefore expect surviving subhalos in SIDM to preferentially occupy tangential orbits relative to surviving subhalos in CDM. We test this by measuring the orbital anisotropy profile
\begin{equation}
    \beta(r) \equiv 1 - \frac{\sigma_{t}(r)^2}{2\sigma_r(r)^2},
\end{equation}
where $\sigma_r$ and $\sigma_{t}$ denote the radial and tangential velocity dispersion of subhalos, respectively. Note that $\beta< 0$ corresponds to a tangentially biased orbital distribution, $\beta = 0$ corresponds to an isotropic orbital distribution, and the maximum allowed value of $\beta=1$ corresponds to a purely radial orbital distribution.

\begin{figure}[t]
\includegraphics[scale=0.4]{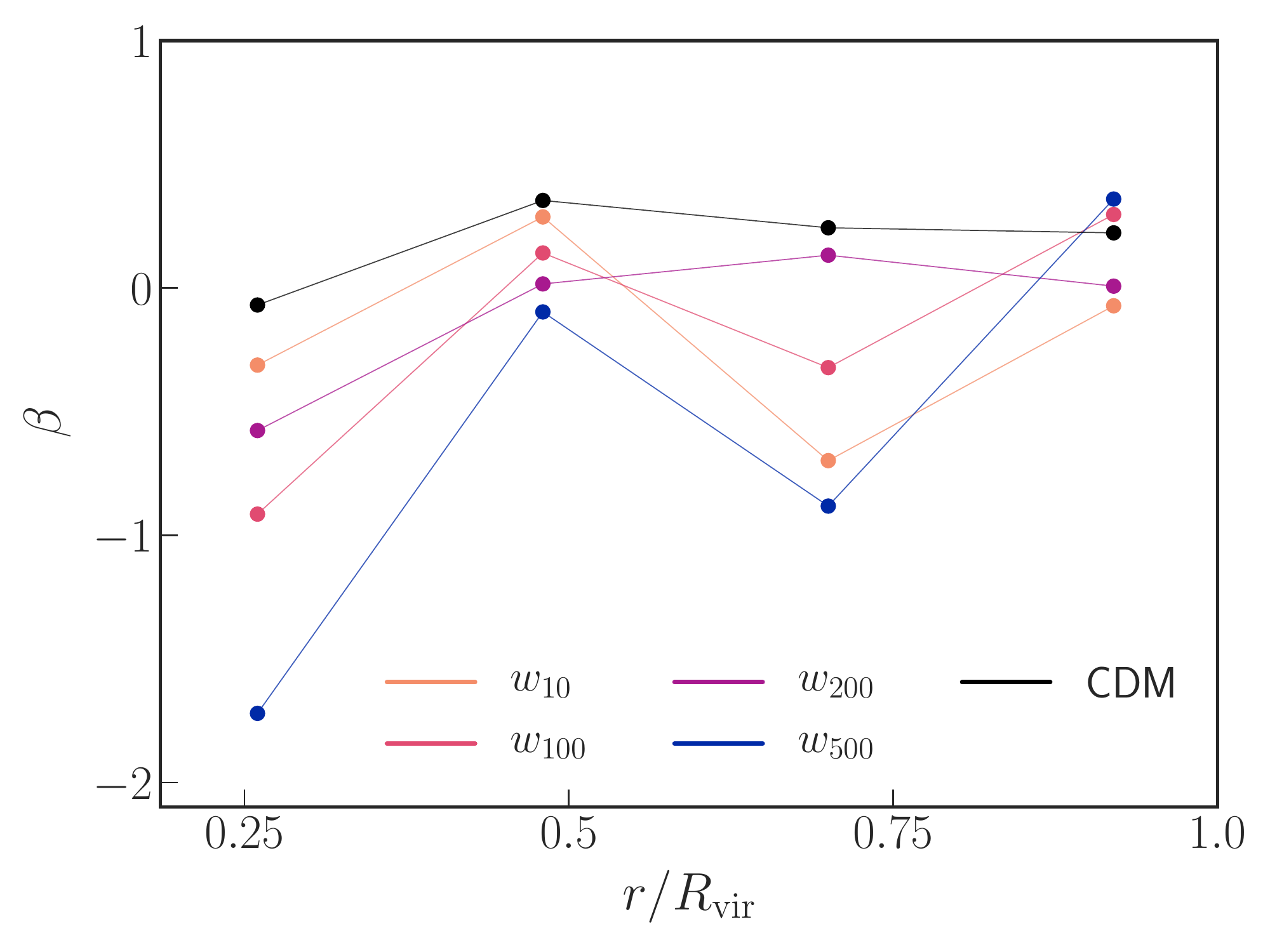}
\caption{Orbital anisotropy profile of subhalos in our SIDM simulations. Surviving subhalos in SIDM model variants with larger values of $w$ occupy tangentially biased orbits relative to CDM. This occurs because a combination of ram pressure and tidal stripping preferentially disrupts subhalos on radial orbits.}
\label{fig:beta}
\end{figure}

We measure $\beta(r)$ by calculating the radial and tangential velocity distributions of subhalos in the frame of the host halo, binning subhalos in distance from the center of the host. The result is shown in Figure \ref{fig:beta}. We find that the orbital anisotropy profile rises toward the outskirts of the host halo in our SIDM simulations, while it is roughly flat in CDM. As expected based on the results in Section \ref{sec:disruption}, surviving subhalos preferentially occupy tangential orbits in our SIDM simulations, and this effect is more significant for model variants with larger values of $w$, for which subhalo disruption via ram pressure plus tidal stripping is more severe.\footnote{The amplitudes of the $\beta(r)$ profiles at small radii are sensitive to the choice of radial binning because there are few resolved subhalos in the inner regions. However, the overall shape of $\beta(r)$ and the trends with $w$ are robust.} We observe a similar trend for anisotropy profiles computed using the DM particles belonging to the host halo in each simulation, although the magnitude of the effect is less severe in this case. An interesting consequence of this effect is that, at fixed $V_{\rm{peak}}$, surviving subhalos in our SIDM simulations tend to be \emph{more} massive on average than their counterparts in CDM, since they are less susceptible to tidal stripping on tangential orbits; however, this difference largely vanishes if subhalos are matched based on their orbital properties. 

These results are interesting in light of recent high-precision measurements of the orbital properties of MW satellites enabled by Gaia (e.g., \citealt{Fritz180500908,Gaia180409381,Simon180410230}). In particular, recent studies of the orbital anisotropy profile find that the typically flat~$\beta(r)$ profiles found in DM--only simulations are in tension with the observed velocity anisotropy profile of MW satellites \citep{Riley181010645}. Baryonic effects, and particularly the tidal influence of the Galactic disk, affect the $\beta(r)$ profile in a similar manner to ram pressure stripping in our SIDM simulations by disrupting subhalos on radial orbits near the center of the host. However, as noted by \cite{Riley181010645}, the changes to $\beta(r)$ due to baryonic physics only resolve the discrepancy with the observed profile if the Galactic disk is sufficiently massive, which in turn depends on the prescription for baryonic feedback and the mass accretion history of the MW system. Adding baryonic subhalo disruption to an orbital distribution that is \emph{already} tangentially biased owing to self-interactions may result in even more tangential bias than observed in the inner regions of the MW, potentially yielding an upper limit on the SIDM cross section at $\sim 200\ \rm{km\ s}^{-1}$.

Thus, surviving subhalos in SIDM occupy tangentially biased orbits relative to subhalos in CDM, leaving a systematic imprint on the orbital anisotropy profile.

\subsubsection{Subhalo Profiles}
\label{sec:subhalo_profiles}

Because the momentum transfer cross section in our SIDM model scales as $v^{-4}$ above the characteristic velocity scale $w$, self-interactions at low relative velocities are strictly more efficient at transferring momentum than interactions at high relative velocities. Thus, we expect the density profiles of low-mass subhalos in our SIDM simulations to be impacted by self-interactions. In contrast to subhalo disruption, it is not clear \emph{a priori} which SIDM model variants will most significantly affect subhalo profiles, since both self-interactions within subhalos and self-interactions with the host halo potentially impact subhalo density profiles.

\begin{figure*}[t]
\centering 
\includegraphics[scale=0.36]{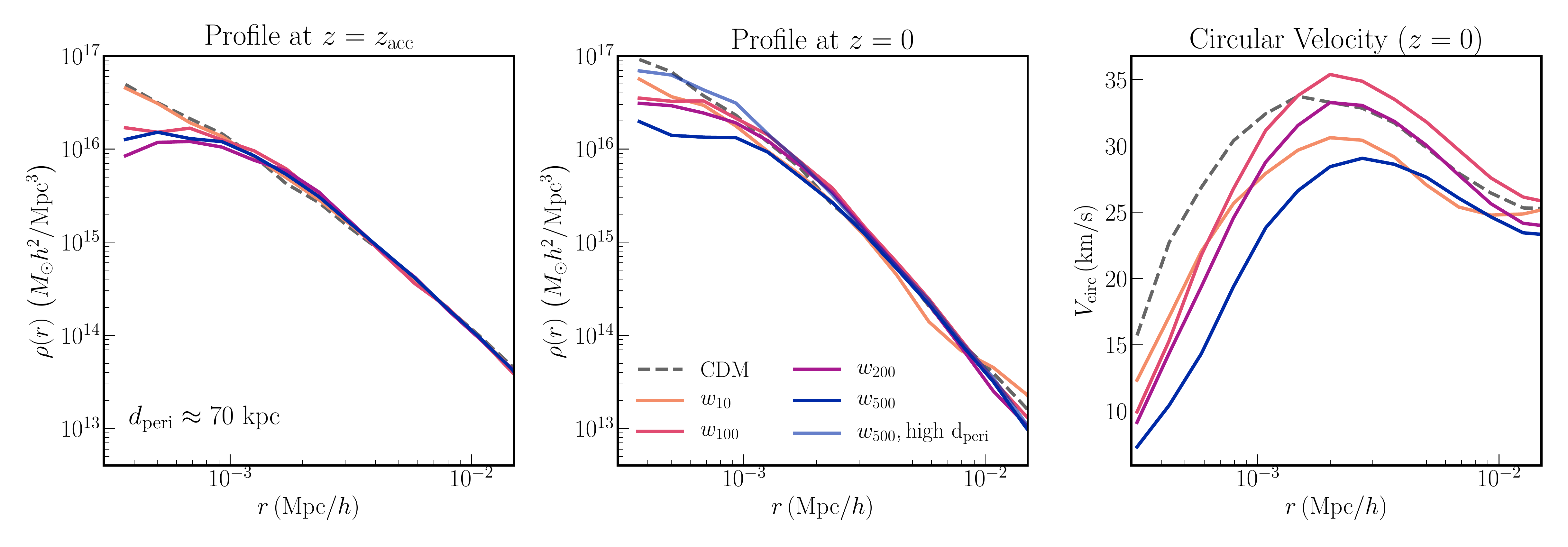}
\caption{DM profiles of a matched set of surviving subhalos in our CDM and SIDM simulations. Density profiles defined by the initial set of bound particles are shown for the same subhalo with $V_{\rm{peak}}\approx 40\ \rm{km\ s}^{-1}$ and $z_{\rm{acc}}\approx 1$ at the time of accretion onto the host (left panel) and at $z=0$ (middle panel). Subhalos in model variants with larger values of $w$ have lower-amplitude, flatter inner density profiles owing to self-interactions with the host halo. The light-blue line in the middle panel shows the density profile for a subhalo with similar $V_{\rm{peak}}$ and $z_{\rm{acc}}$ but with a \emph{large} pericentric distance ($d_{\rm{peri}}\approx 160\ \rm{kpc}$). The discrepancy between this profile and that of the corresponding low-pericenter subhalo demonstrates that the impact of self-interactions on subhalo density profiles depends sensitively on their orbital properties. The right panel shows the corresponding circular velocity profile for each subhalo at $z=0$.}
\label{fig:subhalo_profile}
\end{figure*}

To investigate the impact of self-interactions on subhalo density profiles, we select representative subhalos in our CDM simulation, and we find the corresponding subhalos in our SIDM simulations by matching on $V_{\rm{peak}}$, accretion time, and pericentric distance. We then select the particles in the initial phase-space region associated with these matched subhalos and we track the evolution of their density profiles until $z=0$. Figure \ref{fig:subhalo_profile} shows the results of this procedure for a particular set of subhalos matched to a CDM subhalo with $V_{\rm{peak}}\approx 40\ \rm{km\ s}^{-1}$ and accretion time $z_{\rm{acc}}\approx 1$ that passes close to the center of the host ($d_{\rm{peri}}\approx 70\ \rm{kpc}$). We show the density profiles of these matched subhalos at the time of accretion onto the host and at $z=0$; we also show the density profile of a subhalo in $w_{500}$ with similar $V_{\rm{peak}}$ and $z_{\rm{acc}}$ that does \emph{not} pass close to the center of its host ($d_{\rm{peri}}\approx 160\ \rm{kpc}$).

We find that subhalos in SIDM model variants with larger values of $w$ have lower inner densities and are more cored, i.e., their inner density profiles are flatter as a function of distance from the center of the subhalo than in CDM. We interpret this as a consequence of ram pressure stripping; even though self-interactions within subhalos are more effective for SIDM models with \emph{smaller} values of $w$, interactions with the host at large relative velocities significantly alter the inner profiles of subhalos in model variants with \emph{larger} values of~$w$. As indicated by Figure \ref{fig:subhalo_profile}, we find that---at fixed $V_{\rm{peak}}$ and $z_{\rm{acc}}$---subhalos with closer pericentric passages to the center of the host are significantly more cored. Detailed orbital modeling of satellites in MW-mass systems is therefore crucial in order to interpret the inferred DM density profiles of their subhalos.

We find that the circular velocity profiles of subhalos with close pericentric passages are significantly altered relative to CDM, as expected based on the changes to their density profiles and consistent with previous findings \citep{Vogelsberger12015892,Zavala12116426,Robles170607514,Robles190301469,Fitts181111791,Sameie190407872}. This has implications for SIDM solutions to the diversity and ``too big to fail'' problems concerning MW satellites \citep{Kahlhoefer190410539,Zavala190409998}, although baryonic feedback mechanisms, including heating from supernova feedback, can also reduce and flatten subhalos' central density profiles \citep{Pontzen11060499,Brooks12095394,Creasey161203903,SantosSantos170604202,Read180806634}. However, if cores are created by stellar feedback, then subhalos' inner density profiles are expected to be correlated with their galaxies' star formation histories (e.g., \citealt{Read180806634}). Meanwhile, in SIDM, we expect the inner amplitude and flatness of subhalos' density profiles to correlate most directly with their orbital histories, which can be encapsulated by their infall times and orbital eccentricities.

Thus, surviving subhalos in SIDM models with larger values of $w$ have lower-amplitude, flatter density profiles relative to their CDM counterparts. The magnitude of this effect depends on subhalos' orbital properties such as their infall time and distance of closest approach to the center of the host, and this effect is more significant for larger values of $w$.

\section{Challenges} 
\label{sec:caveats}

We now discuss several caveats and systematics associated with our analysis and main results.

\subsection{Numerical Effects}

Various halo-finding algorithms, including {\sc Subfind}, {\sc AHF}, and {\sc Rockstar} have been used to identify and track halos in cosmological SIDM simulations. Several authors have thoroughly compared these algorithms in the context of CDM simulations (e.g., \citealt{Knebe11040949,Srisawat13073577}). However, the pros and cons of different halo finders and merger tree algorithms are largely unknown for SIDM simulations, and a comprehensive comparison study is crucial in order to go beyond the statistical halo matching technique employed in this paper. Such a study would also be relevant for robust halo finding in hydrodynamic simulations, where baryonic effects like supernova feedback can substantially alter density profiles relative to CDM.

Meanwhile, resolution effects (including artificial disruption; \citealt{VandenBosch171105276,VandenBosch180105427}) are always important to mitigate when analyzing substructure in cosmological simulations, particularly if the abundance of objects near the resolution limit is important. Thus, a detailed study of the underlying mechanisms and numerical stability of subhalo disruption in the presence of self-interactions using extremely high-resolution simulations is another important avenue for future work.

\subsection{Sample Variance}

Because we have simulated a fixed realization of an MW system for various SIDM model variants, we have clearly not sampled a cosmologically representative range of host halo mass accretion histories. Doing so will add scatter to the prediction for the amount of subhalo disruption in SIDM relative to CDM because the number of infalling subhalos and their properties depend on the mass accretion history of the host (e.g., \citealt{Mao150302637}). However, the main physical trends we report, and particularly the correlation between the severity of subhalo disruption and the SIDM cross section evaluated at the host halo velocity scale, should not be affected by marginalizing over mass accretion histories. We plan to test this explicitly by running a suite of zoom-in SIDM simulations. When analyzing MW satellites specifically, certain features of the MW system, such as the properties of the Large Magellanic Cloud system and major accretion events inferred from Gaia data, may lessen the impact of this uncertainty by constraining the allowed range of assembly histories (e.g., \citealt*{PaperII}).

\subsection{Baryonic Effects}

Our analysis is meant to provide insights into the basic physical mechanisms that shape the properties of the MW system in the presence of DM self-interactions. Although we do not include baryonic effects in our simulations, a number of authors have studied SIDM effects on MW and dwarf galaxy scales in the presence of baryons \citep{Kaplinghat13116524,Fry150100497,Robles170607514,Robles190301469,Elbert160908626,Sameie180109682,Fitts181111791}. These studies have revealed that including baryons impacts both the host halo and subhalos in SIDM-only simulations. We now discuss these effects in light of our findings.

\subsubsection{Changes to Density Profiles}

One of the key predictions of the above studies is that the presence of baryons changes SIDM host halo profiles. This occurs because SIDM dynamically responds to the total (i.e., DM-plus-baryonic) gravitational potential (e.g., \citealt{Kaplinghat13116524,Sameie180109682,Robles190301469}). This response results in a cuspier host that is nearly indistinguishable from the host halo in a CDM-plus-baryon simulation. Because this process makes the host cuspier, the subhalo disruption effects reported in this paper should be viewed as lower limits, since both tidal effects and ram pressure stripping will be enhanced in the presence of baryons owing to the host's cuspier inner density profile and increased velocity dispersion. As discussed by \cite{Kahlhoefer190410539}, we do not expect the change to the host halo's density profile to significantly alter surviving subhalo populations, since the host's profile only changes within the few inner kiloparsecs, where very few subhalos survive in our SIDM simulations (see, e.g., Figures \ref{fig:dperi_hist} and \ref{fig:vsidm}). However, including the Galactic disk significantly enhances the likelihood of disruption for subhalos with $d_{\rm{peri}}\lesssim 20\ \rm{kpc}$ (e.g., \citealt{Garrison-Kimmel170103792}).

Unlike the host halo, we do not expect the density profiles of subhalos in our SIDM simulations to change appreciably in the presence of baryons, although they may become more cored owing to tidal interactions with the cuspier host halo and with the Galactic disk. In particular, faint satellites have extremely low stellar masses, which makes both heating from supernova feedback and adiabatic contraction of the DM profile negligible (e.g., \citealt{Elbert14121477,Elbert160908626}). In other words, the presence of a central dwarf galaxy has little effect on the DM profile of its subhalo.

\subsubsection{Accelerated Core Collapse}

By choosing SIDM model variants with reasonably small self-interaction cross sections at all but the lowest velocity scales, we have explicitly avoided studying gravothermal core collapse \citep{Balberg0110561,Ahn0412169,Koda11013097}. Of particular relevance for subhalos orbiting the MW, recent analyses demonstrate that the timescale for core collapse can be significantly shortened owing to tidal stripping by the Galactic disk \citep{Kahlhoefer190410539,Nishikawa190100499}. This accelerated core collapse mechanism has been used to argue that the survival of dense satellites near the center of the MW, and particularly the apparent anticorrelation between satellites' pericentric distances and their inferred central DM densities, is a unique signature of SIDM \citep{Kaplinghat190404939}.

We have chosen to study models in which core collapse is not likely to occur over the timescale of our simulations because this generally requires a very large SIDM cross section, at least at low velocities (e.g., $\sigma_T/m_\chi\gtrsim 3\ \rm{cm^2\ g}^{-1}$ at $v\sim 30\ \rm{km\ s}^{-1}$; \citealt{Kahlhoefer190410539,Nishikawa190100499}). This makes the effect somewhat model dependent because such large cross section amplitudes are likely ruled out by MW satellite abundances for velocity-independent scattering. Furthermore, it is important to study the region of parameter space for velocity-dependent SIDM models that does not result in core-collapsed MW subhalos, since it is not clear whether core collapse is required by the current data.

\section{Comparison to Previous Studies}
\label{subsec:comparison}

Comparing our results to previous studies of SIDM effects on MW-mass systems is not straightforward. For clarity, we limit this discussion to a comparison of results for subhalo abundances. The following theoretical systematics should be kept in mind throughout the discussion in this section:
\begin{enumerate}
\item \emph{SIDM cross section (amplitude and velocity dependence)}: Authors have studied a variety of velocity-dependent SIDM models with different cross section amplitudes. We comment on case-by-case comparisons with our model below.
\item \emph{SIDM implementation}: Most SIDM implementations in the literature are functionally identical to ours, with the exception of \cite{Vogelsberger12015892} (and, by extension, \citealt{Zavala12116426} and \citealt{Dooley160308919}). In particular, we determine whether a given DM particle scatters by calculating the interaction probability with all neighboring particles within a sphere of radius equal to the softening length, while \cite{Vogelsberger12015892} calculate interaction probabilities using the $k=38$ nearest neighbors of each particle.
\item \emph{Halo finding}: Various halo-finding algorithms have been used to analyze SIDM simulations of MW-mass systems.
\end{enumerate}

Our findings are in reasonable agreement with \cite{Vogelsberger12015892}. These authors claim that a velocity-independent SIDM model with $\sigma_T/m_\chi=10\ \rm{cm^2\ g}^{-1}$ (i.e., our $w_{500}$ model with a $10$ times higher cross section) is ruled out by the abundance and density profiles of bright MW satellites. These authors find that subhalo disruption in this velocity-independent model is driven by evaporation due to subhalo--host halo interactions, similar to our conclusion that ram pressure stripping plus tidal effects drive subhalo disruption in $w_{500}$. \cite{Vogelsberger12015892} also find that two velocity-dependent SIDM models, both of which are similar to our $w_{10}$ model but with slightly larger momentum transfer cross sections at the host halo velocity scale, are allowed by the MW satellite data. By visual inspection, it appears that subhalo abundances are suppressed by $\sim 20\%$ relative to CDM in these models, with the largest differences at low subhalo masses. The amount of subhalo disruption in our $w_{10}$ model is consistent with this result (see, e.g., Figure \ref{fig:vsidm}), although we find that the suppression of subhalo abundance relative to CDM is a much weaker function of subhalo mass, possibly due to our use of a matched subhalo sample.

\cite{Rocha12083025} find that the abundance of MW subhalos in SIDM models with velocity-independent cross sections of $\sigma/m_\chi = 0.1$ and $1\ \rm{cm^2\ g}^{-1}$ is very similar to that in CDM. However, they find that $\sim 20\%$ fewer subhalos survive near the inner regions of their host halo~($r\lesssim0.5R_{\rm{vir}}$) for $\sigma/m_\chi = 1\ \rm{cm^2\ g}^{-1}$, corresponding to~$\sigma_T/m_\chi = 0.5\ \rm{cm^2\ g}^{-1}$. Our $w_{100}$ model, in which $\sim 35\%$ of subhalos are disrupted relative to CDM, has a similar momentum transfer cross section at the MW-mass host halo velocity scale. Thus, our results are roughly consistent with \cite{Rocha12083025}, although further investigation into halo finder differences is needed to assess whether the remaining $\sim 15\%$ discrepancy is due to the velocity dependence of our $w_{100}$ model. Interestingly, \cite{Rocha12083025} find that subhalo abundances for~$\sigma/m_\chi = 0.1\ \rm{cm^2\ g}^{-1}$ are identical to CDM at low $V_{\rm{max}}$, and that there are actually \emph{more} high-$V_{\rm{max}}$ subhalos than CDM in this case. Although we cannot compare our results to their~$0.1\ \rm{cm^2\ g}^{-1}$ model directly, we similarly find that the high-$V_{\rm{max}}$ tail of the surviving subhalo distribution is enhanced in our SIDM simulations relative to CDM, which might be due to the fact that surviving subhalos occupy tangential orbits on which tidal stripping is less severe.

\cite{Zavala12116426} simulate velocity-independent SIDM models with $\sigma_T/m_\chi = 0.1$, $1$, and $10\ \rm{cm^2\ g}^{-1}$, as well as two velocity-dependent models that lie between our $w_{10}$ and $w_{100}$ models at the MW-mass host halo velocity scale, but which rise more steeply at low relative velocities. They find that the only model that leads to a difference in substructure abundance relative to CDM is $\sigma_T/m_\chi = 10\ \rm{cm^2\ g}^{-1}$, in stark contrast to our finding that subhalo disruption is significant in all of our model variants. This may be explained by the difference in SIDM implementation, since interactions between subhalo and host halo particles are less likely in the \cite{Zavala12116426} implementation than in ours. In particular, our implementation allows interactions between any DM particles within a softening-length-sized sphere, while the $k$-nearest neighbors technique used in \cite{Zavala12116426} does not. Although \cite{Zavala12116426} use the same SIDM implementation as \cite{Vogelsberger12015892}, we speculate that there is less of a discrepancy between our results and those of \cite{Vogelsberger12015892} because these authors only study models in which the cross section is either negligible or extremely large at the host halo velocity scale.

We note that \cite{Dooley160308919} study stellar stripping using the same set of SIDM models and implementation as \cite{Zavala12116426}. These authors reiterate that substructure is only affected in the most extreme SIDM model, although they remark that subhalos with close pericentric passages in any of the models typically evaporate within~$\sim 6\ \rm{Gyr}$, which is qualitatively consistent with our findings.

Finally, although they focus on the impact of adding a Galactic disk to SIDM simulations, \cite{Robles190301469} simulate MW host halos without disks in CDM and SIDM. The subhalo $V_{\rm{max}}$ functions and radial distributions shown in \cite{Robles190301469} suggest that $\sim 60\%$ of subhalos survive in their~$\sigma_T/m_\chi = 1\ \rm{cm^2\ g}^{-1}$ model relative to CDM, which is in reasonable agreement with our findings. Again, this effect is only noticeable at low $V_{\rm{max}}$. Like the results in \cite{Rocha12083025}, and consistent with our findings, the SIDM-only model in \cite{Robles190301469} shows an \emph{enhancement} in the abundance of high-$V_{\rm{max}}$ subhalos relative to CDM.

\section{Prospects for SIDM Constraints}
\label{subsec:prospects}

The effects reported in this work can be incorporated in models of DM substructure in MW-mass systems in order to constrain the SIDM cross section at relatively low velocity scales. For example, several authors have recently proposed forward-modeling frameworks in which various DM properties can be constrained based on the abundance of observed MW satellites (e.g., \citealt{Jethwa161207834,Nadler190410000,Nadler180905542}). These constraints are set by the fact that deviations from CDM yield a suppression in the abundance of the low-mass subhalos that are inferred to host faint satellite galaxies. Incorporating our finding that subhalo disruption is driven by the self-interaction cross section at the host halo velocity scale will therefore yield an upper limit on the SIDM cross section at $\sim 200\ \rm{km\ s}^{-1}$. The precise value of this limit is difficult to forecast given the theoretical uncertainties discussed above; thus, we plan to carry out this study comprehensively in future work. Meanwhile, gaps and perturbations in stellar streams---which are also sensitive to the abundance, radial distribution, and density profiles of both surviving and disrupted subhalos in the MW---can be used to place complementary constraints (e.g., \citealt{Bovy151200452}).

Our results suggest that placing an upper limit on the SIDM cross section at velocity scales \emph{below}~$\sim 200\ \rm{km\ s}^{-1}$ by probing subhalo abundances is challenging, or at least highly model dependent, since the abundance of surviving subhalos is not very sensitive to the SIDM cross section below the velocity scale of the host. However, if observations of the inner density profiles of the faintest MW satellites unambiguously favor cored or cuspy inner DM density profiles, it may be possible to place a stringent limit on $\sigma_T/m_\chi$ at low relative velocities (e.g., $\sim 10\ \rm{km\ s}^{-1}$). \cite{Read180506934} claim that the cuspy halo profile inferred for the MW satellite Draco places an upper bound on the velocity-independent SIDM cross section of $\sigma/m_\chi \lesssim 0.6\ \rm{cm^2\ g}^{-1}$ ($\sigma_T/m_\chi \lesssim 0.3\ \rm{cm^2\ g}^{-1}$) at $99\%$ confidence. However, as we have argued, modulations to subhalos' density profiles in the presence of self-interactions are highly dependent on their orbital histories; in addition, it is not clear how the constraint in \cite{Read180506934} translates to the velocity-dependent SIDM model considered here. We also note that the stellar profiles of surviving satellites can evolve differently depending on whether they occupy a cored or cuspy halo (e.g., \citealt{Errani150104968}).

High-resolution spectroscopy on giant segmented mirror telescopes (e.g., \citealt{Simon190304743}), along with improvements in mass profile modeling techniques (e.g., \citealt{Genina191109124,Read180806634,Lazar190708841}), will increase the precision of density profile measurements and constraints. However, observational systematics associated with inferring the underlying DM density profiles (e.g., \citealt{Pineda160207690,Oman171202562}), and degeneracies with baryonic coring mechanisms (e.g., \citealt{Read180806634}), will likely make a \emph{detection} of DM self-interactions at low relative velocities challenging. On a positive note, our results demonstrate that observables that cover a range of velocity scales related to MW-mass systems---such as the inferred abundances and density profiles of low-mass subhalos---are highly complementary. In particular, combining such observables informs the velocity dependence of the SIDM cross section, which is a key facet of many well-motivated SIDM models.

\section{Conclusions}
\label{sec:conclusion}

Given increasingly precise constraints on the abundance and internal properties of DM substructure enabled by observations of satellite galaxies, strongly lensed systems, and stellar streams, a detailed examination of DM models at the edge of allowed parameter space is crucial. SIDM is particularly interesting in this context, since constraints on the velocity-dependent self-interaction cross section at relative velocities that are small compared to those typical for galaxy clusters provide an important handle on particle physics in the dark sector. In this paper, we studied the phenomenology of a generic, velocity-dependent SIDM model in the context of the DM host halo and subhalos of MW-mass systems. Due to its velocity dependence, the effects of our SIDM model are inherently scale dependent. Thus, observations of MW-mass halos and their subhalos provide a range of velocity scales with which to probe DM self-interactions.

We have demonstrated that the characteristic velocity above which momentum transfer due to self-interactions becomes inefficient, which is related to the DM and mediator masses in our SIDM model, has a variety of phenomenological consequences for MW-mass host halos and their subhalos. Our main findings are as follows:
\begin{enumerate}
\item In the absence of baryons, the DM distribution in the inner regions of the host is cored and thermalized owing to self-interactions (Figure \ref{fig:host_properties}). This effect is stronger for larger values of the cross section at the host halo velocity scale, corresponding to our model variants with larger values of $w$.
\item The initial assembly of subhalos in our SIDM simulations, quantified by $V_{\rm{peak}}$ and the time at which $V_{\rm{peak}}$ is achieved, is statistically similar to that in CDM (Figure \ref{fig:vpeak_hist}).
\item Subhalos are mildly affected by self-interactions before infall into the host, leading to lower maximum circular velocity values and moderately cored density profiles at infall relative to subhalos in CDM (Figures \ref{fig:vpeak_hist} and \ref{fig:subhalo_profile}).
\item A significant fraction of the subhalos that survive in CDM are not found in SIDM owing to extra mass loss from ram pressure stripping caused by self-interactions with the host halo, which makes subhalos more susceptible to tidal disruption. This effect is more severe for models with larger values of $w$ (Table \ref{tab:sidm_sims}, Figures \ref{fig:dperi_hist} and \ref{fig:vsidm}).
\item Surviving subhalos in SIDM occupy tangentially biased orbits relative to those in CDM, causing a systematic trend in the orbital anisotropy profile that is more significant for models with larger values of $w$ (Figure \ref{fig:beta}).
\item Surviving subhalos in SIDM models with larger values of $w$ are less dense and more cored than surviving subhalos in CDM, and the magnitude of this effect depends sensitively on orbital properties such as infall time and pericentric distance, with more severe coring for subhalos that pass closer to the center of their host (Figure \ref{fig:subhalo_profile}).
\end{enumerate}

Given these findings, we plan to carry out a comprehensive study of these effects for a simulated MW-like host halo that includes a realistic Large Magellanic Cloud analog system, which has recently been used to fit the full-sky MW satellite luminosity function (\citealt*{PaperII}). This analysis will allow us to constrain the SIDM cross section at the MW host halo velocity scale ($\sim 200\ \rm{km\ s}^{-1}$) using the abundance, surface brightness distribution, and radial distribution of observed MW satellites.

We emphasize that self-interactions also affect the orbital distribution and density profiles of subhalos in MW-mass systems. Comparing these quantities to data in a forward-modeling approach will help break degeneracies among SIDM models that are not ruled out by subhalo abundances alone, and will therefore inform the velocity dependence of the SIDM cross section. For example, surviving subhalos in our SIDM simulations preferentially occupy tangential orbits, and this prediction can be compared to the measured orbital distribution of MW satellites using proper-motion measurements from Gaia and its future data releases. Meanwhile, surviving subhalos in SIDM exhibit cored density profiles, and the strength of this effect is a function of both microphysical SIDM parameters and the orbital history of each subhalo.

We expect that combining high-precision spectroscopic and proper-motion measurements of satellite galaxies with analyses of strongly lensed systems and perturbations in stellar streams will test these predictions, providing a new window into velocity-dependent DM self-interactions.

\acknowledgments

We thank Manoj Kaplinghat, Annika Peter, and the anonymous referee for comments on the manuscript, and we thank Neal Dalal, Xiaolong Du, and Carton Zeng for useful discussions. 
This research received support from the National Science Foundation (NSF) under grant No.\ NSF DGE-1656518 through the NSF Graduate Research Fellowship received by E.O.N., by the U.S. Department of Energy contract to SLAC No. DE-AC02-76SF00515, and by Stanford University.
Y.-Y.M.\ is supported by NASA through the NASA Hubble Fellowship grant No.\ HST-HF2-51441.001 awarded by the Space Telescope Science Institute, which is operated by the Association of Universities for Research in Astronomy, Inc., under NASA contract NAS5-26555. We thank the LSST Dark Matter Group (\href{https://lsstdarkmatter.github.io/}{lsstdarkmatter.github.io}) for feedback at workshops supported by the LSSTC Enabling Science program (grant No.\ 2017-11).

This research made use of computational resources at SLAC National Accelerator Laboratory, a U.S.\ Department of Energy Office, and the Sherlock
cluster at the Stanford Research Computing Center (SRCC); the authors are thankful for the support of the SLAC and SRCC computing teams.  This research made extensive use of \https{arXiv.org} and NASA's Astrophysics Data System for bibliographic information.

\bibliographystyle{yahapj2}
\bibliography{references}


\begin{figure*}[t]
\includegraphics[scale=0.35]{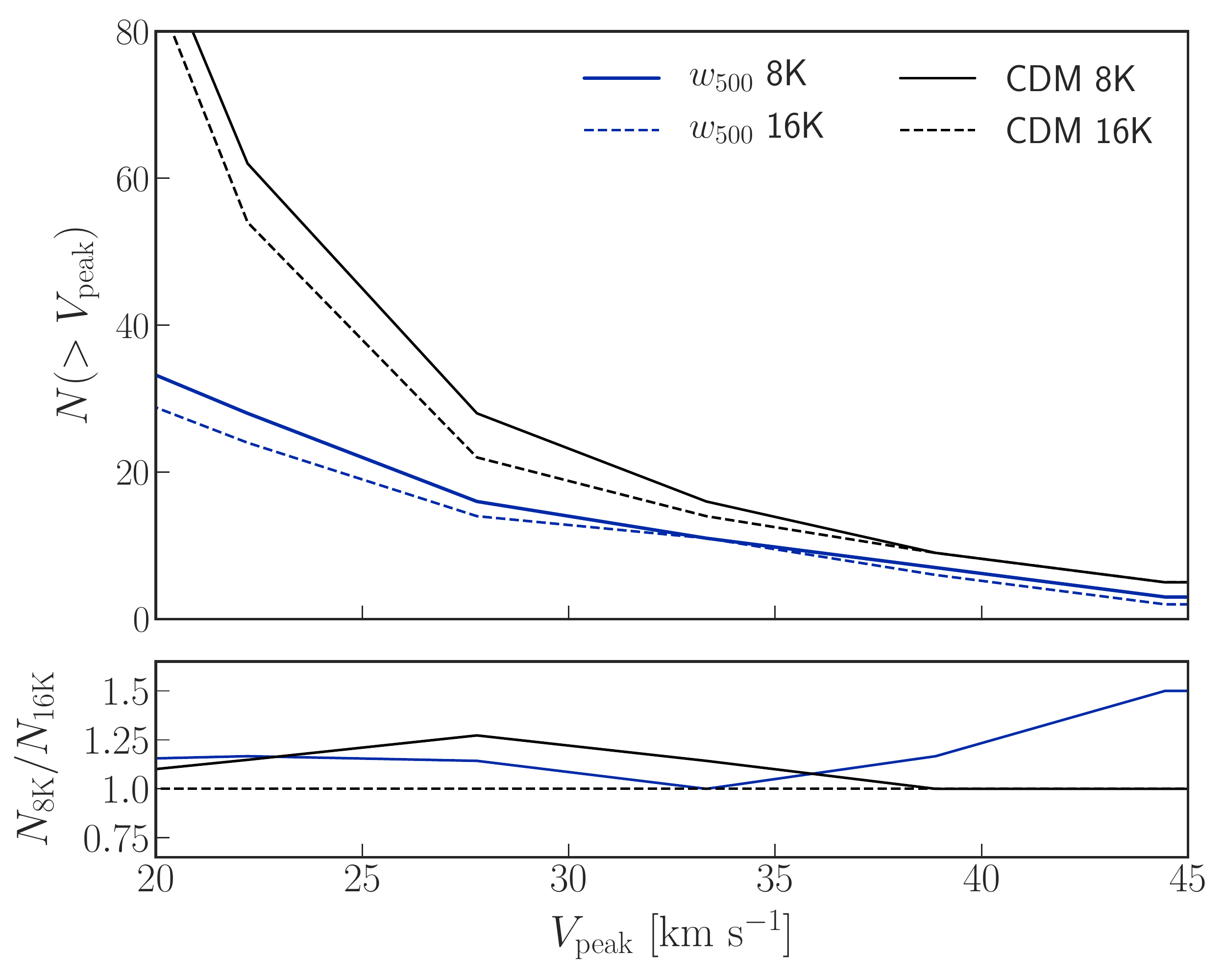}
\includegraphics[scale=0.35]{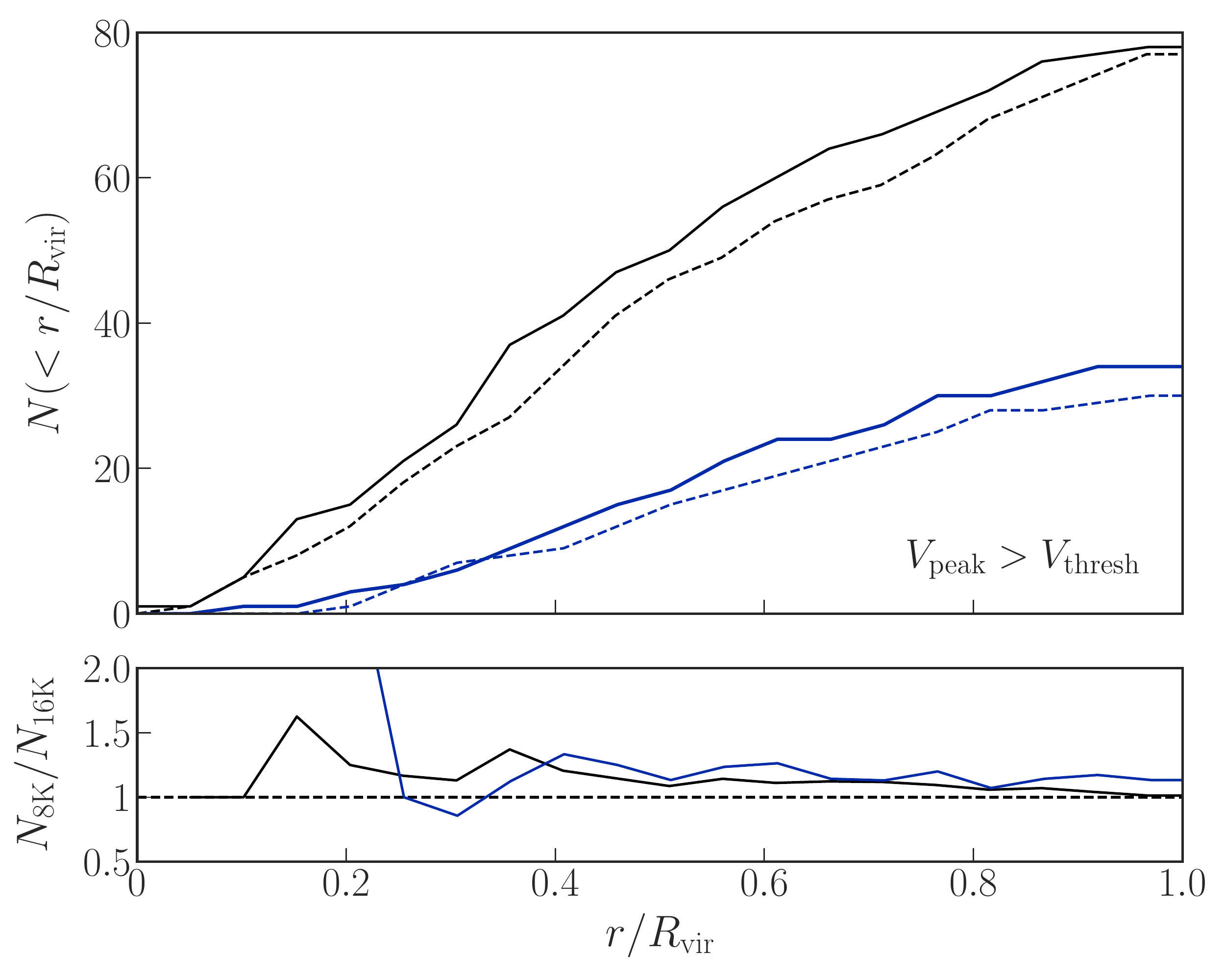}
\caption{Simulation resolution study. Left panel: peak velocity functions of subhalos for our high-resolution CDM and $w_{500}$ resimulations (labeled ``16K''), compared to those from our fiducial simulations (labeled ``8K''). Right panel: corresponding radial subhalo distributions in units of the host halo virial radius in each simulation.}
\label{fig:res_test}
\end{figure*}

\appendix

\section{Simulation Resolution Tests}
\label{appendixa}

To test for convergence, we rerun our CDM and $w_{500}$ simulations at higher resolution. In particular, these resimulations are run with a $4.0\times 10^{4}\ M_{\rm \odot}\ h^{-1}$ high-resolution particle mass and an $85\ \text{pc}\ h^{-1}$ minimum softening length, corresponding to a factor of eight increase in mass resolution and a factor of two decrease in softening length relative to our fiducial simulations.

In Figure \ref{fig:res_test}, we compare the subhalo $V_{\rm{peak}}$ functions and radial distributions from these high-resolution resimulations to our fiducial results. We find fairly good agreement, at the $\sim 15\%$ level, between the standard and high-resolution results for subhalos above our fiducial $V_{\rm{peak}}$ thresholds. Interestingly, we find that there are \emph{more} resolved subhalos above the relevant $V_{\rm{peak}}$ threshold in the lower-resolution simulations, which might be due to the fact that early interactions within subhalos are better resolved in the higher-resolution simulations, leading to more efficient coring. However,  we note that there are also more resolved subhalos in the lower-resolution CDM simulation, so this also might indicate that the $V_{\rm{peak}}$ distributions are slightly different in the fiducial and high-resolution simulations; we have not attempted to match subhalo populations precisely for this comparison, since the level of agreement is reasonable. These findings suggest that artificial subhalo disruption is not a large effect relative to the physical disruption (or stripping below the resolution limit) of subhalos in our SIDM simulations. Thus, we conclude that our main results, including the amount of subhalo disruption in our SIDM simulations, are not highly sensitive to the resolution threshold of our fiducial simulations.

\section{Subhalo Resolution Threshold}
\label{appendixb}

In our fiducial analysis, we employed a conservative $V_{\rm{peak}}$ resolution cut of $20\ \rm{km\ s}^{-1}$ for our CDM simulation, and we matched the number of surviving-plus-disrupted subhalos in CDM and each SIDM model variant using variable $V_{\rm{peak}}$ thresholds, denoted by $V_{\rm{thresh}}$. In this appendix, we show that this statistical subhalo matching method is necessary given the scatter in the~$V_{\rm{peak}}$ distributions measured by our halo finder. We then demonstrate that our matched subhalo populations are well converged given our fiducial choice of $V_{\rm{thresh}}=20\ \rm{km\ s}^{-1}$ in CDM. Finally, we summarize our main results for a lower $V_{\rm{peak}}$ threshold.

\subsection{Subhalo Abundance for a Fixed $V_{\rm{peak}}$ Threshold}

\begin{figure*}[t]
\includegraphics[scale=0.4]{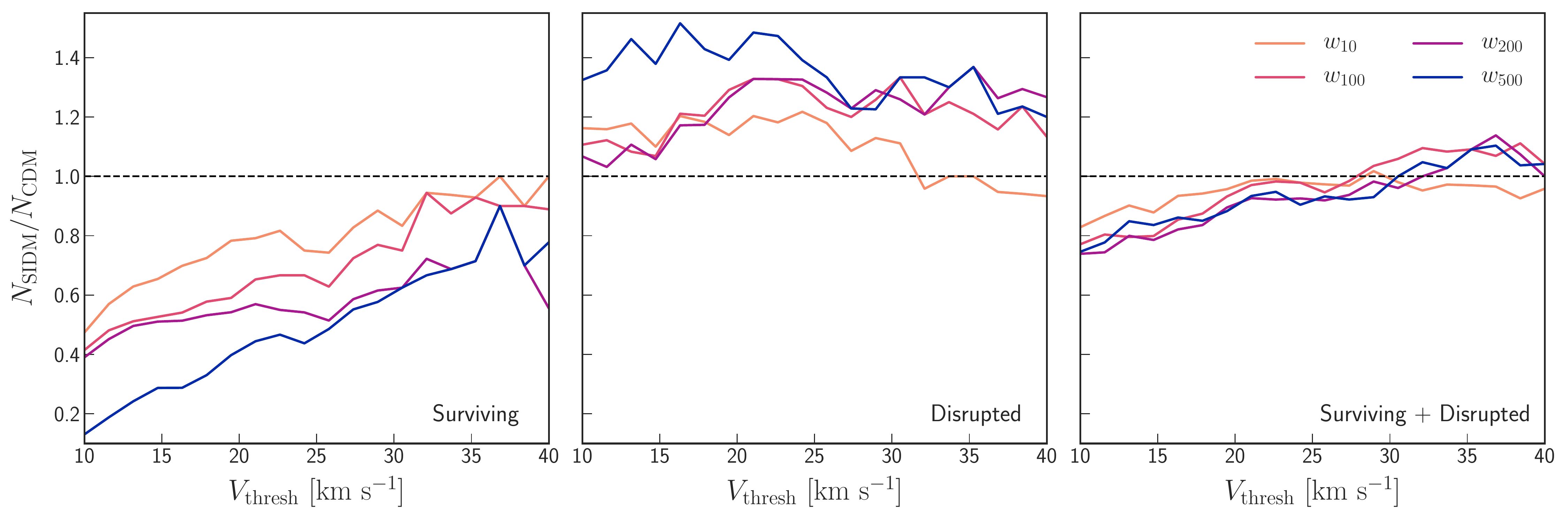}
\caption{Subhalo resolution threshold study. Left panel: total number of surviving subhalos in each SIDM model variant above a fixed $V_{\rm{peak}}$ threshold divided by the corresponding number of surviving subhalos in our CDM simulation. Middle panel: same as the left panel, but for the number of disrupted subhalos. Right panel: same as the previous panels, but for the number of surviving-plus-disrupted subhalos.}
\label{fig:vthresh}
\end{figure*}

Figure~\ref{fig:vthresh} shows the number of surviving, disrupted, and surviving-plus-disrupted subhalos in SIDM relative to that in CDM if the \emph{same} value of $V_{\rm{thresh}}$ is used for all of the simulations. The scatter among the SIDM model variants in all three panels of Figure~\ref{fig:vthresh} strongly suggests that a variable $V_{\rm{peak}}$ threshold must be adopted in order to match subhalo populations in our CDM and SIDM simulations. Physically, this is reasonable because of the differences in subhalo assembly on an object-by-object basis caused by early self-interactions described in Section \ref{sec:pre-infall-evolution}; however, the differences could also be driven by halo finder issues. Note that the scatter among our SIDM model variants increases for lower values of $V_{\rm{thresh}}$, making precise subhalo population matching more difficult near the CDM resolution limit of $V_{\rm{peak}}\approx 10\ \rm{km\ s}^{-1}$.

\subsection{Choice of Fiducial $V_{\rm{peak}}$ Threshold}

We observe that the \emph{total} number of subhalos is stable at the $\sim 10\%$ level for $V_{\rm{thresh}}\gtrsim 20\ \rm{km\ s}^{-1}$, lending confidence to our fiducial subhalo resolution threshold. The residual differences relative to CDM for large values of $V_{\rm{thresh}}$ are likely due to a combination of physical effects (i.e., true differences in subhalo assembly in the presence of self-interactions) and numerical artifacts (e.g., uncertainties inherent to using halo finders optimized for CDM on our SIDM simulations, or artificial subhalo disruption). Disentangling these effects is beyond the scope of this paper; however, as noted in Section \ref{subsec:comparison}, this $\sim 10\%$ effect must be accounted for as a theoretical systematic until the cause of these differences is well understood.

\subsection{Severity of Subhalo Disruption for a Less Conservative $V_{\rm{peak}}$ Threshold}

Finally, we estimate the severity of subhalo disruption in our SIDM simulations for a lower $V_{\rm{peak}}$ threshold. In general, we expect enhanced disruption for subhalos with $V_{\rm{peak}}\lesssim 20\ \rm{km\ s}^{-1}$ because these systems are even more susceptible to tidal disruption than those studied in our fiducial analysis. Indeed, we observe that the fraction of surviving subhalos in SIDM falls off sharply at low $V_{\rm{peak}}$ thresholds, particularly for $V_{\rm{thresh}}\lesssim 15\ \rm{km\ s}^{-1}$, and we have confirmed that this behavior persists in our higher-resolution $w_{500}$ resimulation. We also note that, in SIDM models with large self-interaction cross sections at low relative velocities (e.g., $w_{10}$), self-interactions \emph{within} subhalos can significantly affect their density profiles, again making them more susceptible to tidal disruption. There is a hint of this effect in the left panel of Figure \ref{fig:vthresh}, where we observe the steepest downturn in the number of surviving subhalos relative to CDM for $w_{10}$ near the CDM resolution limit of $V_{\rm{peak}}\approx 10\ \rm{km\ s}^{-1}$.

It is unclear whether the downturn in the fraction of surviving subhalos in our SIDM simulations is a consequence of physical disruption, artificial disruption, and/or halo finder issues in this regime. However, for completeness, we calculate the fraction of surviving subhalos in SIDM relative to that in CDM using our statistical subhalo matching procedure for $V_{\rm{thresh}}=15\ \rm{km\ s}^{-1}$ in CDM. This choice yields $V_{\rm{thresh}}=[14.36,13.93,13.84,14.07]\ \rm{km\ s}^{-1}$ and a surviving subhalo fraction of $N_{\rm{SIDM}}/N_{\rm{CDM}}=[0.72,0.63,0.65,0.31]$ for $[w_{10},w_{100},w_{200},w_{500}]$, respectively. These surviving fractions are nearly identical to our fiducial results for $w_{100}$ and $w_{200}$, and they are consistent at the $\sim 10\%$ ($30\%$) level for $w_{10}$ ($w_{500}$). In addition, the $V_{\rm{peak}}$ distributions of surviving-plus-disrupted subhalos remain consistent among the simulations using these $V_{\rm{thresh}}$ values.

\section{Subhalos near the Host Center}
\label{appendixc}

As noted in Section \ref{sec:population_stats}, there is a curious ``spike'' in the inner radial distribution of surviving subhalos for $w_{10}$ and $w_{200}$, shown in the right panel of Figure \ref{fig:vsidm}. The spike is particularly significant for $w_{10}$, where the number of subhalos in one of the inner radial bins is \emph{enhanced} by a factor of two with respect to CDM. We emphasize that these spikes are not statistically significant: they only correspond to one or two additional subhalos near the center of the host in the SIDM simulations. Given the increased efficiency of tidal disruption in SIDM, it might seem surprising that these subhalos survive. While the survival (relative to our other SIDM model variants) and sinking (relative to CDM) of these objects are plausible consequences of self-interactions, we cannot distinguish the presence of these subhalos at small radii from statistical fluctuations. In particular, it is possible that pericentric passages at $z\approx 0$ determined by the orbital phases of the subhalos in $w_{10}$ and $w_{200}$ happen to align more closely with the $z=0$ snapshot than for subhalos in CDM or our other SIDM model variants. On the other hand, if this behavior persists in a larger suite of SIDM simulations, it may be a physical consequence of increased drag due to self-interactions.

\end{document}